\documentclass[10pt,a4paper]{article}
\usepackage{arxiv}
\usepackage{color}
\usepackage{amssymb,amsmath,graphicx,cite}
\usepackage{bm}
\usepackage{braket}
\usepackage{makeidx}
\usepackage{epsfig}
\usepackage{array}
\usepackage{multirow}
\usepackage[utf8]{inputenc} 
\usepackage[T1]{fontenc}    
\usepackage{hyperref}       
\usepackage{url}            
\usepackage{booktabs}       
\usepackage{amsfonts}       
\usepackage{nicefrac}       
\usepackage{microtype}      
\usepackage{lipsum}	
\usepackage{listofsymbols}
\usepackage{moreverb}
\usepackage{listings} 
\title{Nonequilibrium Phase Transition in a 2D Ferromagnetic Spins with Effective Interactions}
\author{Dagne Wordofa Tola$^{1,2}$   and  
Mulugeta Bekele$^{1,}$\thanks{Corresponding Author}\\
$^{1}$Department of Physics, Addis Ababa University, POB~1176, Addis Ababa, Ethiopia\\  
$^{2}$Department of Physics, Dire Dawa University, POB~1362, Dire Dawa, Ethiopia\\  
Email: $^{*}$\textit{mulugetabekele1@gmail.com} (MB)}

\begin{document}
\maketitle
\tableofcontents
\setcounter{tocdepth}{1}
\addtocontents{toc}{\protect\setcounter{tocdepth}{1}}
\begin{abstract}
The study of nonequilibrium steady-state (NESS) in the Ising model offers rich insights into the properties of complex systems far from equilibrium. This paper explores the nature of NESS phase transitions in two-dimensional (2D) ferromagnetic Ising model on a square lattice under effective interactions using Monte Carlo (MC) algorithms. This requires extensive MC simulations using the modified Metropolis and modified Glauber update rules. The qualification of the modified update rules is characterized by the definition of an effective parameter $h$. For $|h|>1$, it is analytically shown that the nature of the phase transition (including the critical temperature) is independent of $h$. Furthermore, for $-1<h<1$, we study the steady-state properties of phase transitions using numerical methods. Therefore, we performed simulations for different lattice sizes and measured relevant physical quantities. From the data, we determined the numerical results of the transition temperature and relevant critical exponents for various values of $h$ by applying finite-size scaling (FSS). We found that the FSS analysis of the exponents is consistent with the analytical values of the equilibrium 2D Ising model.
\end{abstract}
\section{Introduction}\label{Introduction}
The fundamental principles and standard theory of critical behavior near continuous phase transitions (PTs) in equilibrium systems are currently well understood~\cite{kardar2007lattice, Peters2012,linares2021, Goldenfeld1992}. Ongoing research and exploration have, however, continuously focused on the investigation of PTs between nonequilibrium statistical states~\cite{Cates2000, Derrida2007, Derrida2011, Bertini2015, Godreche2009, Henkel2010,Stinchcombe2001,Odor2004,
Hinrchsen2000,acharyya2011,dickman1999,Kumar2020}. Despite considerable efforts, research problems associated with the classification of nonequilibrium PTs have still not been completely solved. In the equilibrium model, PT is generically represented by singularities in the free energy and its derivatives. Such a singularity causes a discontinuous property of the physical quantities near the critical point. PT is phenomenologically characterized by an {\it order parameter} that is vanishingly zero in the disordered phase and non-vanishing in the ordered phase~\cite{Onsager1944, YangI1952, YangII1952}. Out-of-equilibrium systems exhibit a broad range of universality classes, such as ``kinetic Ising models with competing  dynamics''~\cite{Mallick2023, Gonzalez1987, Dickman1987, Szolnoki2000, Tome1991, Marques1990, Godoy2002, Garrido1989}.

Nonequilibrium PTs are a wide research area that is increasing in most situations that are intrinsically out of equilibrium, in which the standard tools of equilibrium statistical mechanics are usually not applicable. However, fundamental concepts such as criticality and universality have been extended to nonequilibrium prototypes. Nonequilibrium PTs can be classified into several types such as directed percolation, active matter, and self-organized criticality. These are just a few examples, and there are more examples of nonequilibrium PTs. Each type of PT has its typical properties and mathematical formalism, and understanding these transitions is indispensable for studying the broad fields of complex systems across multiple disciplines. Intriguingly, this is far less understood, although classes have been studied over the last few decades. Nonequilibrium PTs are an essential research field because much of their functional nature resides in out-of-equilibrium conditions; for example, quantum annealing is used for some contemporary quantum computers for which the Ising model is directly relevant. In addition to describing several magnetic systems, the Ising model can be used to examine the critical behaviors in different gases, alloys, glasses, and liquid helium mixtures~\cite{lipowski2022ising}. However, the focus of this study is specific to the application of the Ising model to phases and PTs in magnetic systems.  More specifically, we consider a paradigmatic example of ferromagnetic and paramagnetic phases and the transitions between these phases; therefore, we use the language of a 2D ferromagnetic spin system. Despite many recent efforts on related works, the nature of nonequilibrium PTs in Ising ferromagnets with effective interactions has not yet been studied.%

In a related paper~\cite{Kumar2020}, the Monte Carlo (MC) method using a modified Metropolis and a modified Glauber algorithm was proposed to study the nature of nonequilibrium PT in the 2D Ising model, including the order of the PT, as well as its universality class. More specifically, nonequilibrium PT, where the detailed balance condition (DBC) is not fulfilled and the system reaches a nonequilibrium steady state (NESS) that is not described by Boltzmann statistics, was addressed using modified Metropolis algorithms. In the model studied there, “activity” was introduced by modifying the update rule in a way that makes the probability of occurrence of spin flips higher than that in the equilibrium 2D Ising model. The opposite case called a "persistent Ising Model" in which spin flips are less likely to occur than in the equilibrium model was also studied. The name persistence model was used because the modified rule increases the persistence time of the spin configurations. Recently, \cite{tola2023machine} implemented a supervised machine learning adaptive approach based on convolutional neural networks to predict the critical temperature of the nonequilibrium transition from the paramagnetic to ferromagnetic state in a 2D Ising model on a square lattice, in which the MC simulation relies on a previous study~\cite{Kumar2020}.  Although the modified update rule was less accurate for the ``persistent'' regime, the agreement between the two numerical methods was excellent. In this study, we propose a strategy in which an effective Hamiltonian suffices for the nonequilibrium description of the mathematical model to study the steady-state nature. Interestingly, the equilibrium 2D Ising model can be solved precisely~\cite{Onsager1944}. This could be helpful in estimating the transition temperature in the case of the NESS. We employ mean field approximation (MFA), which is a prominent method to exhibit a qualitative picture of the problem. We use the MC method~\cite{berg2004, LandauBinder2014} to provide numerical results.

The remainder of this paper is organized as follows. In Section~\ref{S2_Methods}, we define the model considered in this study and provide a concise description of the methods. Section~\ref{S3_MFA} discusses the problem using MFA, including a qualitative demonstration. The numerical results are presented and discussed in detail in Section~\ref{S4_Results}. Finally, we conclude with a summary of the main findings in Section~\ref{S5_Summary}. 

\section{Methods}\label{S2_Methods}
\subsection{Ferromagnetic Ising model and the MC method}\label{subsection:Ising}
Consider a 2D ferromagnetic Ising model on the lattice of linear size $L$ comprising $V = L^{d}$ sites where $d$ denotes the number of spatial dimensions. Each site on the lattice contains a single spin pointing upwards or downwards, and the total number of spins is equal to the size of the system, $N = V$. Let a Hamiltonian describing PTs in Ising effective interactions consists of two terms, $ \mathcal{H} = \mathcal{H}_{0} + \mathcal{H}_{1}$ where $\mathcal{H}_{0}$ refers to the standard equilibrium Ising Hamiltonian and $\mathcal{H}_{1}$ is the interaction part other than $\mathcal{H}_{0}$. Therefore
\begin{equation}\label{Eq:201}
    \mathcal{H} =- J \sum_{\langle i,j \rangle} S_{i}  S_{j}-\sum_{ij' \neq \langle ij\rangle}J_{i j'}  S_{i} S_{j'}  ,
\end{equation}
where $S_i$ is the value of the spin variable at site $i= \{1, \cdots, N \}$ that can be either $\pm 1$(up or down), indices of sites $\langle i,j \rangle$ denotes the sum over all nearest-neighbor pairs~\cite{berg2004, LandauBinder2014, Nishimori2011Numerical} while $ij'\neq \langle ij\rangle$ prompts sum over others except the nearest-neighbors while the third sum is over all $N$ sites. 
The statistical average of an observable $\langle \mathcal{O} \rangle$ derived from canonical  partition sum $Z$ is
\begin{equation}\label{Eq:202}
\langle \mathcal{O} \rangle =   \frac{1}{Z} \sum \mathcal{O} \exp\left[- \frac{\mathcal{H} }{k_{B}T}  \right],
\end{equation}
using $Z = \sum \exp\left[-  \beta \mathcal{H} \right]$ and $\beta = 1/(k_{B}T)$ where  $T$ is temperature and  $k_{B}$ is Boltzmann constant. If distinction of the spin as an operator is not required, the total energy of the Ising model can be given as  $E = E^{0} + E^{1}$ where $ E^{0} = \langle \mathcal{H}_{0} \rangle$ and $ E^{1}  = \langle \mathcal{H}_{1}\rangle $. We define the \textit{nearest-neighbor equilibrium} Ising energy as usual,
\begin{equation}\label{Eq:203}
 E^{0}  = - J  \sum_{\langle ij \rangle}  S_i  S_j .
\end{equation}
Let us now introduce a modified formalism$-$perhaps conceptually `simple' approach$-$of total energy calculation by considering other interactions into account. The primary objective of this manifestation is to express $E^{1}$ in terms of $E^{0}$~\ref{Eq:203} such that  $E^{1}:= h E^{0}$, where $h$ is an effective parameter to be determined bearing on its qualification to establishing the model's NESS nature (see~\ref{S22} and Section~\ref{S3_MFA}). Thus,
\begin{equation}\label{Eq:204}
E(h)   = - J(1+h)  \sum_{\langle ij \rangle}  S_i  S_j.
\end{equation}
This Equation~\eqref{Eq:204} is similar to the usual nearest neighbor model with a coupling constant $J$ replaced by $J(1+h)$. However, this is not our goal. We emphasize a setup in which the term $J(1+h)$~\ref{Eq:204} helps in the qualification of an update rule. Depending on an effective update rule and the choice of $h$, this term may describe NESS, as will become clear later in~\ref{S22}. Henceforward, we use $ E^{0} = E(h = 0)$ to represent the usual energy (without the effective interaction), and $E = E(h \neq 0)$ represent the energy with the effective interaction for the sake of simplicity. We also keep this notation consistent for others, such as temperature $T^{0}=T(h=0)$ and $T=T(h\neq 0)$. Note that the unit of the temperature is related to that of $J/k_B$ where the ferromagnetic energy scale $J>0$ stands for the coupling strength of each spin to its nearest neighbors. To simplify the notation, we will now set $k_B = 1$, hence $T:= T/J$ is dimensionless.

Based on the postulates of equilibrium ($h=0$) statistical physics for the system in contact with a heat bath (thermal reservoir) at a given temperature, each of the spin configurations $\{S\}$ happens with a probability $\mathcal{P}^{0} \propto \exp[- \beta^{0} E_{\{S\}}^{0}]$. The transition temperature of the equilibrium Ising model with $d-$number of spatial dimensions was derived~\cite{Onsager1944} to be
\begin{equation}\label{Eq:205}
 T_{c}^{0} =zJ/\ln(3+2\sqrt{2}) ,
\end{equation}
where $z=2d$ represents the possible number of the nearest-neighbor spins. We aim to test and show how also this scenario works for the case $h \neq 0$. At this point, it is natural to ask some questions: { \it What a setup and choice of the effective parameter $h$ determines the stochastic dynamical behavior?}
In other words: { \it How does modifying an update rule and changing $h$ affect the stochastic dynamical system?  If so, does it also affect the nature of PT?} As a matter of fact, the postulate is applicable also in the case of $h\neq 0$ for those values of $h$ in which the equilibrium condition holds for dynamic processes approaching thermal equilibrium. Here, stochastic dynamical processes play a crucial role in the equilibrium models, such as $N$-particle systems in contact with the heat bath at $T=T(h)$. At a limit of an asymptotically long time, this system approaches a statistically stationary state in which it gets through certain configuration $\{S\}$ according to a well-defined probability distribution. An explicit form of the above partition sum $Z_{N}(h)$ is now given as $Z =\sum\exp[-(1+h)\beta E_{\{S\}}^{0}]$, where the sum runs over all possible $\{S\}$ configurations. Consequently, $Z$ reads  
\begin{equation}\label{Eq:206}
  Z =    \sum_{\{S\}}\left(\exp\left[-\beta E^{0}\right]\frac{}{}\right)^{(1+h)},
\end{equation}
where $\exp\left[-\beta E^{0}\right]$ is the well known Gibs-Boltzmann distribution. Certain macroscopic quantities of interest can be derived from~\eqref{Eq:206}.

For the model considered in this paper, we show that the system encounters a second order PT at temperature $T_{c}$ for $h\ge -1 $.  However, while $T_{c}$  depends on $h$  for  $|h|\le 1$, it does not depend on $h$  for $h>1$. The system magnetizes for  $T < T_{c}$ where the resulting state is ordered state. The system is considered to be  in the disordered state for $T > T_{c}$.   An order parameter usually defined as an average magnetization per site, $m=\langle M\rangle/N$ where
\begin{equation}\label{Eq:207}
\begin{array}{c}M =\left| \sum_{i} S_i\right|,\end{array}
\end{equation}
and it quantitatively distinguishes the two phases realized by the system. That is, zero $m$ refers to the state in which the orientation of spins is disordered, and non-zero $m$ corresponds to the state in which it is in a preferred direction. Determining $T_{c}$ and analyzing the nature of transition will be competent tasks of this paper. Primarily, however, we need to define the parameter $h$, including a justification of its effectiveness.
\paragraph{Monte Carlo simulation method:}
Consider a system that generates stochastic spin flips when in contact with a thermal reservoir~\cite{Glauber1963}. In the standard Ising model, the system achieves thermal equilibrium after a sufficiently long time, and the description of the steady-state distribution is subject to the Boltzmann distribution. This permits one to define transition rates and calculate the flipping probabilities. Nonequilibrium PT is discussed with emphasis on general features such as the role of breaking DBC in generating effective interactions~\cite{racz2002nonequilibrium}. The DB is an adequate condition$-$it is not a necessary condition to ensure equilibration\cite{Cates2000}. In this study, we construct a simple model that violates DBC and causes the system to come out of equilibrium, depending on the nature of the effective interaction in consideration. In agreement with the discussion in related paper~\cite{Kumar2020}, there is a situation in which the system shows a disorder-order transition that is not similar to the usual equilibrium PT. As each spin flips from time to time due to the influence of thermal fluctuation, we use the master equation (ME) to develop this idea in terms of stochastic changes in its configurations. The intrinsic stochastic dynamics of the system allow us to compute its thermodynamic properties. Most of our knowledge about equilibrium PT can be extended to the nonequilibrium case as well.

Denoting the spin configuration before the flip $\texttt{b}= \{S_{1}, ...,\pm S_{i}, ...,S_{N}\}$, the system's state is described according to probability theory that the state has a configuration $\texttt{b}$ at time $t$, with probability $\mathcal{P}_{b}(h,t)$. Assuming that the configuration after the flip will be $\texttt{a}= \{S_{1}, ..., \mp S_{i}, ..., S_{N}\}$, the transition changes  from state $\texttt{b}$ to $\texttt{a}$ with rate of transition probability $W(b\rightarrow a)$ in a couple of time interval $\Delta t$ \cite{Nishimori2011Numerical}. Consequently, the probability that the configuration of system being in state  $\texttt{b}$ decreases by $W(b\rightarrow a)\Delta t \mathcal{P}_{b}(h,t)$, since the system was in  $\texttt{b}$ with probability $\mathcal{P}_{b}(h,t)$ and then has changed to  $\texttt{a}$ with rate of transition probability $W(b \rightarrow a)$. Correspondingly, the probability that it is in state  $\texttt{b}$ would be increased by 
$W(a\rightarrow b)\Delta t \mathcal{P}_{a}(h,t)$.
We shall use a discrete form of ME and define the net change of probability per unit $\Delta t$ as follows:
\begin{equation}\label{Eq:208}
\frac{\Delta \mathcal{P}_{b}(h,t)}{\Delta t}=\sum_{a\neq b} W( a\rightarrow b)\mathcal{P}_{a}(h,t)- \sum_{a\neq b} W( b\rightarrow a)\mathcal{P}_{b}(h,t),
\end{equation}
where $\Delta  \mathcal{P}_{b}(h,t)= \mathcal{P}_{b}(h,t+\Delta t) - \mathcal{P}_{b}(h,t)$.
This ME~\eqref{Eq:208} is an implicit idea of Markov processes where the characteristic change of $\mathcal{P}_{b}(h,t)$ is completely described as an effective probability distribution at time $t$ while $h$ is fixed. Though a proposition that the dynamics are generated by heat baths secure that the rates~\eqref{Eq:208} satisfy DB at some control parameters, we intended to deal with rates violating DB. In sufficiently long-time limit $t\to \infty$, a stochastic dynamical system approaches a statistically stationary state where it evolves through certain configuration according to a distinct probability distribution that does not change with time and $\Delta\mathcal{P}_{b}(h,t\to \infty) \approx 0 $. As a result, the ME is reduced to DBC, 
\[ W( b\rightarrow a)\mathcal{P}_{b}(h)  =  W(a\rightarrow b)\mathcal{P}_{a}(h) ,\]
\begin{equation}\label{Eq:209}
 \mathcal{R}(h) = \frac{W(b\rightarrow a)}{W(a\rightarrow b)}  \equiv \frac{\mathcal{P}_{a}}{\mathcal{P}_{b}}  ,
\end{equation}
where $\mathcal{R}$ denotes a \emph{ratio} of the transition probabilities (or called flipping probability), and $\mathcal{R}(h=0)$ is equivalent to the well-known Boltzmann weighting. However, the distribution of stochastic dynamical system approaching thermal equilibrium may not necessarily be the Gibbs-Boltzmann distribution~\cite{Nishimori2011Numerical}. Hence it will be customary for $h\neq 0$ to choose an appropriate weight in MC simulations that $\mathcal{P}_{b,a}(h) = \exp[-\beta E_{b,a}(h)]/Z $, and get%
\begin{equation}\label{Eq:210}
\mathcal{P}_{a}/\mathcal{P}_{b} = \exp[-\beta \Delta E ].
\end{equation}
where $ \Delta E = E_{\rm{a}} - E_{\rm{b}}$ is the energy change due to the transition from a present state \verb"b" (`before the flip') to a new state \verb"a" (`after the flip').
Using Equation~\eqref{Eq:204},
\[\Delta E = J(1+h)\sum \left(S_i^{\rm{b}} - S_i^{\rm{a}}\right)S_j = 2J(1+ h)S_{i}\sum_{j}  S_{j},\]
where the spin $ S_i$ before and after the flip has opposite sign ($ S_i^{\rm{b}} = -   S_i^{\rm{a}}\equiv S_i$), while the nearest neighbors remain the same ($ S_{j}^{\rm{a}} =    S_{j}^{\rm{b}}\equiv  S_{j}$).
Therefore,
\begin{equation}\label{Eq:211}
 \Delta E =   (1+ h) \Delta E^{0}  ,  
 \end{equation}
where  $ \Delta E^{0}$ is defined as $\Delta E^{0} = 2 J  S_{i}\sum_{j}  S_{j}$ with  $- \Delta E_{\rm{max}}^{0} \leq \Delta E^{0} \leq \Delta E_{\rm{max}}^{0}$. Using a simple version of the Ising (spin~$1/2$) like Hamiltonian~\cite{Salinas2001} described in Equation~\eqref{Eq:201} in which  $\Delta E_{\rm{max}}^{0} = 2 zJ$ and each spin is linked to other nearest neighbors.  For a square lattice Ising  in equilibrium ($h=0$), we have $z=4$ and $ \Delta E^{0} $ can take a set of values
\begin{equation}\label{Eq:212}
\Delta E^{0} =  \{-8, -4, 0, 4, 8\} .
\end{equation}
Equation~\eqref{Eq:210} helps to describe the two most commonly used transition rates namely the Metropolis~\cite{Metropolis1953} update rule
\begin{equation}\label{Eq:213}
W =  \underbrace{\texttt{MIN} \left[1, \exp\left[ -\beta  \Delta E \right]\frac{}{}\right]}_{\rm (A\ref{Eq:213})}
 \;  \Rightarrow   \; \underbrace{\texttt{MIN} \left[1, \exp\left[-\beta  d E \right]\frac{}{}\right]}_{\rm (B\ref{Eq:213})} ,
\end{equation}
and the Glauber~\cite{Glauber1963,Janke2007,Susuki1968} update rule
\begin{equation}\label{Eq:214}
W =  \underbrace{\frac{1}{2} \left( 1- \tanh \left[ \frac{\beta \Delta E}{2} \right] \right)}_{\rm (A\ref{Eq:214})} \;  \Rightarrow  \; 
\underbrace{\frac{1}{2} \left( 1- \tanh \left[ \frac{\beta d E}{2} \right]\right)}_{\rm (B\ref{Eq:214})}.
\end{equation}
The arrow declares that an update rule is modified through replacing $\Delta E$ by $dE$, where
\begin{equation}\label{Eq:215}
\begin{array}{c}
d E =  \Delta E^{0} +  h |\Delta E^{0}|\rm{, \; \; \; \; }  \Delta E^{0 } = \{-8, -4, 0, 4, 8\} .
\end{array}
\end{equation}
Therefore, replacing $\Delta E$ in (\verb"A"\ref{Eq:213}) and (\verb"A"\ref{Eq:214}) by this effective $d E$~\eqref{Eq:215}, we get the modified update rules (\verb"B"\ref{Eq:213}) and (\verb"B"\ref{Eq:214}).  For the sake of clarity, we mention model \verb"A" and model \verb"B",  where \verb"B" is the modified version of the well-known  \verb"A". Accordingly, we call (\verb"A"\ref{Eq:213}) and (\verb"B"\ref{Eq:213}) in case of the Metropolis update rule, whereas (\verb"A"\ref{Eq:214}) and (\verb"B"\ref{Eq:214}) in the case of the Glauber dynamics. Before we study the nature of PTs, it is essential to check whether or not \verb"B" breaks the DBC. For results reported in Section~\ref{S4_Results}, we manage the simulation of model \verb"B" for the specified system on a square lattice of linear size  $20<L<160$ by applying periodic boundary conditions in each direction. We start the simulations from a high value of $T$ in the range and set an initial state to be random configurations. For each $T$ in the temperature list and a specified value of $h$, we choose a random site $i$ of spin $S_{i}$ from $L\times L$ lattice. A detail of our MC method and examples are available from~\cite{tola899} in which we use a Python-based jupyter-notebook for demonstration purposes.

Notice that the two dynamics~\eqref{Eq:213} and~\eqref{Eq:214} may exhibit different properties near the critical temperature. The former tends to be more efficient at exhibiting phase space and exploring critical phenomena accurately. The latter may explore slower dynamics and tend to be less effective at capturing critical phenomena efficiently. Therefore, in some cases, the critical properties determined with the Glauber dynamics might not be similar to those obtained with the Metropolis algorithm. Accordingly, it is often possible to choose the update rule based on the specific properties of the physical system under investigation as well as the phenomena of interest.
\subsection{Occurrence of phase transition using the modified update rules}\label{S22}
\begin{figure}[hbpt]
     \centering
\includegraphics[width=0.9 \columnwidth]{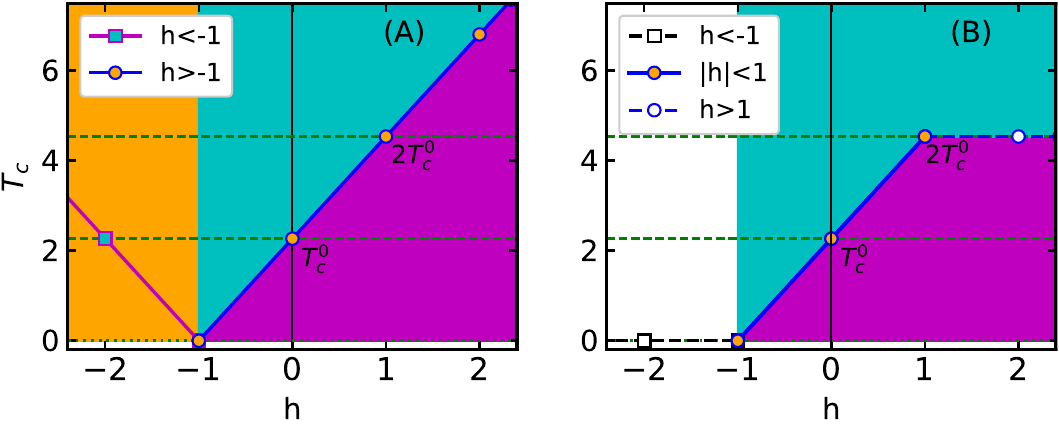}
   \caption{\small{A plot of $T_{c}$ vs $h$ for Equation~\eqref{Eq:222} A and~\eqref{Eq:224} B where left (A) and righ (B) models are corresponding to $\Delta E$~\eqref{Eq:211} and $dE$~\eqref{Eq:215}, respectively. The horizontal dashed lines represent  $T_{c}^{0}$ and  $2T_{c}^{0}$ where $ T_c^{0} = 2/\ln(1+\sqrt{2})$.}} \label{fig1}
\end{figure}

In this paper, we implement the two modified update rules~(\verb"B"\ref{Eq:213}) and~(\verb"B"\ref{Eq:214}), as briefly introduced in~\ref{subsection:Ising}. A description of the update rule~\eqref{Eq:213}, is given in~\cite{tola899} (see also~\ref{A:A1}). Here, we make our presentation specific to the Glauber dynamics and discuss the occurrence of PT using Equation~\eqref{Eq:214}, but the same procedure applies to Equation~\eqref{Eq:213}. The update rule~(\verb"A"\ref{Eq:214}) can be rewritten as
\[ W = \frac{1}{2}\left( 1-\tanh \left[\frac{\beta\Delta E}{2}\right] \right)  {, \; \; \;  } W' =\frac{1}{2}\left( 1+ \tanh \left[\frac{\beta\Delta E}{2}\right] \right)  \rm{,} \]
where $ W = W(S_{i}\to -S_{i})$ and $ W' = W(-S_{i}\to S_{i})$. Consequently, the ratio~\eqref{Eq:209}$-$for Glauber dynamics$-$here reads
\begin{eqnarray}
\mathcal{R}(\verb"A"\ref{Eq:214})& =  \frac{1-\tanh \left[\beta\Delta E/2\right]}{1+\tanh \left[\beta\Delta E/2\right]}\texttt{, \; \; } \Delta E = (1+h)\Delta E^{0}  , \label{Eq:216a}\\
\mathcal{R}(\verb"B"\ref{Eq:214})& = \frac{1-\tanh \left[\beta dE/2\right]}{1+\tanh \left[\beta dE/2\right]}\texttt{, \; \; } dE = \Delta E^{0} + h|\Delta E^{0}|, \label{Eq:216b}
\end{eqnarray}
where \eqref{Eq:216b} is a modified version of~\eqref{Eq:216a} in which $\Delta E$ is replaced by $dE$.
In this work, reasonably we choose Equation~\eqref{Eq:216b} to perform MC simulation~\cite{tola899}.
Make use of Equation~\eqref{Eq:215} into~\eqref{Eq:216b}, we need to consider the following two cases: 
\begin{eqnarray}\label{Eq:217}
\mathcal{R}(\verb"B"\ref{Eq:214}) & =  \frac{1-\tanh \left[\beta\Delta E^{0}(1+h)/2\right]}{1+\tanh \left[\beta\Delta E^{0}(1+ h)/2\right]}, \hbox{\; if \; } \Delta E^{0} > 0  , \\  \label{Eq:218}
  & = \frac{1+\tanh \left[\beta|\Delta E^{0}|(1-h)/2\right]}{1-\tanh \left[\beta|\Delta E^{0}|(1- h)/2\right]},  \hbox{\; if \; } \Delta E^{0} < 0  . 
\end{eqnarray}
Notice that the ratio~\eqref{Eq:217} and~\eqref{Eq:218} each equals a unity at $h=-1$ and at $h=1$, respectively,
\[\mathcal{R}(\verb"B"\ref{Eq:214}, h=-1)  = 1, \hbox{\; if \;} \Delta E^{0} > 0  \texttt{,  and \; }    \mathcal{R}(\verb"B"\ref{Eq:214}, h=1)  = 1,   \hbox{\; if \; } \Delta E^{0} < 0  ,  \]
but
\begin{eqnarray}\label{Eq:219}
\mathcal{R}(\verb"B"\ref{Eq:214}, h= 1) & =  \frac{1-\tanh \left[\beta\Delta E^{0}\right]}{1+\tanh \left[\beta\Delta E^{0}\right]} \equiv \exp[-2 \beta\Delta E^{0}], &  \hbox{if \; } \Delta E^{0} > 0  ,\\ \label{Eq:220}
 \mathcal{R}(\verb"B"\ref{Eq:214}, h=-1) & = \frac{1+\tanh \left[\beta|\Delta E^{0}|\right]}{1-\tanh \left[\beta|\Delta E^{0}|\right]}  \equiv \exp[2 \beta|\Delta E^{0}|] , & \hbox{if \; } \Delta E^{0} < 0  . 
\end{eqnarray}
This means that the modified Glauber satisfies DBC for $h=-1$ and $h=1$.
According to the Glauber update rule for the usual Ising model at temperature $T^{0} =1/\beta^{0}$, the transition probability is,
\begin{equation}\label{Eq:221}
\mathcal{R}(h=0)  =  \frac{1-\tanh \left[\beta^{0}\Delta E^{0}/2\right]} {1+\tanh \left[\beta^{0}\Delta E^{0}/2 \right]} 
\equiv \exp[-\beta^{0}\Delta E^{0} ] .
\end{equation} 
Application of the usual update rule~\eqref{Eq:221}, which satisfies DB at temperature $T^{0}$, generates equilibrium configurations of the Ising model at $T^{0}$. It is straightforward that the update rule ~(\verb"A"\ref{Eq:214}) satisfies DB at {\em effective} temperature $T_{\rm{eff}} = T/(1 + h)$ and generates equilibrium Ising configurations at this temperature. Figure~\Ref{fig1} illustrates  phase diagram of the two models where the left panel and right panel corresponds to model A and model B.
In agreement with Figure~\ref{fig1}(A), the system undergoes a PT when  $T_{\rm{eff}} = T_{c}^{0}$, i.e., at {\em critical} temperature $T_{c} = (1+h)T_{c}^{0}$ where $T_{c}^{0} = 2/\ln(1 + \sqrt{2})$ is the critical temperature of the usual Ising model. Indeed, for both (\verb"A"\ref{Eq:213}) and (\verb"A"\ref{Eq:214}) on the square lattice ($z=4$), genuinely it is obvious from the scaling that the critical temperature~\eqref{Eq:205} at a given $h$ becomes
\begin{equation}\label{Eq:222}
T_{c}(h) =  {2(1+h)}\left/{\ln\left(1+\sqrt{2}\right)}\right. .
\end{equation}
Conversely, for $-1<h<1$, model \verb"B" does not {\emph{uniquely} correspond to an equilibrium system  at a fixed effective temperature. Therefore, MC simulation using the modified update rule (\verb"B"\ref{Eq:214}) generates NESS that can be studied for different values of $h$. We can rewrite Equation~\eqref{Eq:217} and~\eqref{Eq:218} as,
\begin{equation} \label{Eq:223}
\mathcal{R}(\verb"B"\ref{Eq:214}) = \left\{
  \begin{array}{ll}
   \exp \left[-\beta\Delta E^{0}(1+h)\right],   & \hbox{ if } \Delta E^{0}  >  0; \\
   \exp \left[\beta|\Delta E^{0}|(1-h)\right],  & \hbox{ if } \Delta E^{0}  < 0 .
  \end{array}
\right.
\end{equation}
For $-1\le h\le 1$, the mapping between Equation~\eqref{Eq:223} and~\eqref{Eq:221} provides
\[-\beta^{0}\Delta E^{0}    = \left\{
  \begin{array}{ll}
   -\beta\Delta E^{0}(1+h)  & \hbox{if } \Delta E^{0}  >  0; \\
    \beta|\Delta E^{0}|(1-h)  , & \hbox{if \;} \Delta E^{0}  < 0 .
  \end{array}
\right. \]
Therefore, equivalently and in agreement with that of the Metropolis update rule (see~\ref{A:A1} ~\eqref{Eq:A07}), the PT occurs at critical temperature,
\begin{equation}\label{Eq:224}
 T_{c}(h, \Delta E^{0}) = \left\{
  \begin{array}{ll}
   (1+h)T_{c}^{0}  & \hbox{if } \Delta E^{0}  >  0; \\
    (1-h)T_{c}^{0}  , & \hbox{if \;} \Delta E^{0}  < 0 ,
  \end{array}
\right.
\end{equation}
where $T_{c}(h=0)= T_{c}^{0}$ as expected. Using $h=1$ into~\eqref{Eq:224} gives $T_{c}= 2T_{c}^{0}$ for $\Delta E^{0}  >  0$ and  $T_{c}= 0$ for $\Delta E^{0}  <  0$. The opposite is true with $h=-1$. 
For model \verb"B", Equation~\eqref{Eq:222} is possibly useful as long as $-1\le h\le 1$~\eqref{Eq:224}.
However, if $\Delta E^{0}  >  0$, Equation~\eqref{Eq:224} doesn't make sense for $h<-1$. It also doesn't make sense for $h >1$ when $\Delta E^{0}  <  0$. This interpretation is consistent with a qualitative demonstration in Figure~\ref{fig1}(B) in which $T_{c}$ satisfies
\begin{equation}\label{Eq:225}
 T_c (h): = \left\{
  \begin{array}{ll}
   0 < T_{c}(h) <  T_{c}^{0}     & \hbox{for } -1 < h  < 0; \\
  T_{c}^{0} < T_{c}(h) < 2 T_{c}^{0} & \hbox{for }  0 < h  < 1,
  \end{array}
\right.
\end{equation}
where $T_{c}^{0} \approx  2.2692$ and $2T_{c}^{0} \approx  4.5384$ are used.
In some cases, the transition temperature of \verb"B"  may correspond to that of model \verb"A" for values of $-1 \le h\le 1$.  If $h > 1$,  model \verb"B" is equivalent to model \verb"A" at $h=1$. Model \verb"B" is equivalent to \verb"A" at $h=-1$, whereas  model \verb"A" becomes antiferromagnetic for $h < -1$. Specifically, if we consider two  $h$ values ($h  = \pm 0.25$), conveniently we get that $T_{c}(h= +0.25)=5/\ln(3 + 2\sqrt{2}) \approx 2.83648$  and  $T_{c}(h= -0.25)=3/\ln(3 + 2\sqrt{2})\approx 1.70189$ and this is in agreement with numerical results~\ref{S4_Results} (see also Section~\ref{S5_Summary} Table \ref{T1}).
\section{The modified update rules with mean-field theory}\label{S3_MFA}
In this section, we treat the solution of the model under investigation using the mean-field (MF) theory, also known as the Curie-Weiss molecular field approximation (MFA). We derive self-consistent equations (SCEs) based on the Metropolis and Glauber dynamics and demonstrate the results quantitatively and qualitatively. Though MF prediction is quantitatively incorrect, the model’s qualitative behaviour is identical to the standard solution. \paragraph{Metropolis dynamics:} Considering approximate schemes for accounting the interactions between spins, let us begin with the  Metropolis dynamics of a spin flip, $S_i \to - S_i$: 
\begin{equation}\label{Eq:326}
w = \underbrace{\texttt{MIN} \left[1, \exp\left[ -\beta  \Delta E \right]\frac{}{}\right]}_{\rm A}
 \;  \to  \; \underbrace{\texttt{MIN} \left[1, \exp\left[-\beta  d E \right]\frac{}{}\right]}_{\rm B} .
\end{equation} 
This is similar to Equation~\ref{Eq:213} but here we use the MF formalism of A and B in which $ \Delta E^{0}:= 2 J z \bar{m} S_{i} $, where $z=2d$ represents the possible number of neighbors for a given $d$. 
 Therefore, the MF formalism of~\eqref{Eq:211} and~\eqref{Eq:215} here read, respectively
 \begin{eqnarray}\label{Eq:327a} 
\Delta E &= & 2 J z \bar{m} (1+h)S_{i}  , \\ \label{Eq:327b}
    d E  &= &  2 J z \bar{m} (S_{i}+h)  .
\end{eqnarray}
Here in MFA, we assume that each spin interacts with a \verb"kind of magnetic cloud"~\cite{silva2023meanfield} described by the {\it mean} magnetization $\bar{m} = \langle S_{i}\rangle $. 
Then substitute~\eqref{Eq:327b} in~\eqref{Eq:326} to get the (modified) MF rate of transition,
\begin{equation}\label{Eq:328}
w =					\left\{
  \begin{array}{ll}
    \exp\left[ -2 \beta J z \bar{m}(S_{i}+h) \right], & \hbox{}d E  > 0 ; \\
    1, & \hbox{otherwise.}
  \end{array}
\right.
\end{equation}
Incorporating this MFA, let us describe SCE for $\bar{m}$ and solve it numerically. Let the spin variable at site $i$ is $S_{i}=\{s_\bullet, s_\circ\}$ where $s _{\bullet} = 1 $  and $s_{\circ } = -1$. Thus $w_{\bullet \rightarrow  \circ}$ denotes the probability of flipping from $s_{\bullet}$ to $s_{\circ}$ and, similarly, $w_{\circ \rightarrow \bullet}$ denotes the probability of flipping from $s_{\circ}$ to $s_{\bullet}$ in the steady state of the system. By defining $r=r(\beta, \bar{m},h)$ as the ratio of these probabilities, it follows that
\begin{equation}\label{Eq:329}
r = \frac{w_{\bullet \rightarrow  \circ}} {w_{\circ \rightarrow \bullet}} 
\equiv \frac{\mathcal{P}_{\circ}}{\mathcal{P}_{\bullet} }.		
\end{equation}
Consequently, we express MF magnetization $\bar{m}$ in terms of $r$ to get the required SCE,
\[\bar{m}(h) =  \frac{1}{Z} \sum_{ \{s\}=\pm 1}  \mathcal{P}_{\{s\}}S_{i}, 
 \mathrm{ \; where \; \; } Z =  \mathrm{e}^{-\beta E_{\bullet}} +  \mathrm{e}^{-\beta E_{\circ}}, \]%
\begin{equation}\label{Eq:330}
\bar{m}(h)   = \frac{ 1- r(\beta, \bar{m},h)}{1+ r(\beta, \bar{m},h)} .
\end{equation}
This SCE can be solved merely for some values of $h$ for which the DBC is satisfied.
 Proceeding with equilibrium dynamics ($h=0$), $r(\beta, \bar{m}, h=0) = \exp [-2\beta^{0} Jz\bar{m}]$ where here $\beta^{0} = \beta^{0}(\textsc{MF})$ denotes $1/T(h=0)$. Consequently
\[ \bar{m}(h=0)   = \frac{ 1- \exp [-2\beta^{0} J z \bar{m}]}{1 + \exp [-2\beta^{0} J z\bar{m}]} ,\]
\begin{equation}\label{Eq:331}
\bar{m}(h=0)  = \tanh[\beta^{0} J z\bar{m}]\texttt{ (Equilibrium SCE),} 
\end{equation}
which is the well-known SCE for the magnetization and we refer to this~\eqref{Eq:331} as  \verb"original SCE" of the equilibrium Ising model.

Given the modified MF version $d E$\eqref{Eq:327b}, it is important to sort the following two statements: (i) If $d E$ is always {\it nonnegative}, then $r = \exp [-4 \beta  J z \bar{m}]$. 
Equation~\ref{Eq:330}, therefore, satisfies a SCE of MF magnetization,  
\begin{equation}\label{Eq:332}
\bar{m}(h \neq 0)   = \frac{ 1-r}{1+ r} = \tanh[2 \beta J z \bar{m}],
\end{equation}
which is independent of $h$. 
 (ii) If $d E$  is always {\it non-positive} and the ratio $r = 1$, therefore it turn outs that
\begin{equation}\label{Eq:333}
\bar{m}(h \neq 0)   = 0 .
\end{equation}
In contrast to~\eqref{Eq:332} and~\eqref{Eq:333}, the purpose of this paper is mainly intended to the case of $h$-dependent SCE as long as $-1<h<1$ establishing a microscopic irreversibility of the dynamics. We noticed from  SCEs~\eqref{Eq:327b} and~\eqref{Eq:329} that $\bar{m}(|h| > 1)$ is independent of $h$, see ~\cite{tola2024nonequilibrium} for more information. For the purpose of this paper, therefore, we compute $\bar{m}(-1 < h <  1)$ as,
\begin{equation}\label{Eq:334}
\bar{m} = \tanh[\beta J z |\bar{m}|(1+h)], \texttt{(Metropolis SCE).}
\end{equation}
The same SCE~\eqref{Eq:334} can be obtained also using Equation~\ref{Eq:327a} which means that the two different forms~($\Delta E$~\eqref{Eq:327a} and~$dE$~\eqref{Eq:327b}) give the same results for $-1 < h <  1$. 
\paragraph{Glauber dynamics:} To go any further, the \texttt{Glauber SCE} should be the same as that of the \texttt{Metropolis SCE} for  both $\Delta E$\eqref{Eq:327a} and $dE$\eqref{Eq:327b}. Using the former~\eqref{Eq:327a} now we can easily solve the Glauber algorithm with a dynamics of single spin flip $S_{i}  \rightarrow  -S_{i}$ as
\begin{equation}\label{Eq:335}
w = \frac{1}{2}\left( 1-S_i \tanh [\beta  J z \bar{m}(1+h)]\frac{}{}  \right) . 
\end{equation}
This result holds true also for $d E$\eqref{Eq:327b} luckily due to the fact that $ \tanh(a-b) = - \tanh(b-a)$ and, therefore, $ \tanh(S_i+h) = S_i \tanh(1+h)$ with $S_i = \pm 1 $.
 Straightforward to the Metropolis dynamics, this Glauber dynamics gives %
 $ w_{\bullet \rightarrow  \circ} =   \frac{1}{2}\left(1-\tanh [\beta  J z \bar{m}(1+h)]\right)$, and %
 $ w_{ \circ \rightarrow \bullet } = \frac{1}{2}\left(1+\tanh [\beta  J z \bar{m}(1+h)]\right)$. %
The ratio~\eqref{Eq:329} now becomes 
\[ r(\beta, \bar{m},h) = \frac{1 - \tanh [\beta  J z \bar{m}(1+ h)]}{1 + \tanh [\beta  J z \bar{m}(1+ h)]},\] thus, one can find $\bar{m}$ as in Equation~\ref{Eq:330} providing a $\texttt{Glauber SCE}$: $\bar{m}  =  \tanh [\beta  J z \bar{m}(1+ h)]$. The Glauber SCE is the same as the Metropolis SCE~\eqref{Eq:334} in the vicinity $-1<h<1$ where the DBC does not hold. The DBC holds for $h=0$ where $\bar{m}(h = 0 ) = \tanh [\beta^{0}  J z \bar{m}]$, and for $h=1$ where $\bar{m}(h = 1 ) = \tanh [2 \beta  J z \bar{m} ]$. In practice,  numerically the modified form of energy difference for \emph{both} dynamics is therefore,
\begin{equation}\label{Eq:336}
d E = \left\{
  \begin{array}{ll}
  2 J z \bar{m} (S_{i}+h), & \hbox{ if } -1 \leq h \leq 1; \\
  4 J z \bar{m} S_{i} ,  & \hbox{ if } |h| >1,
  \end{array}
\right.   
\end{equation}
with the corresponding definition of magnetization SCEs~\eqref{Eq:334}.
\paragraph{Qualitative demonstration:}
The MF results~\eqref{Eq:334} and~\eqref{Eq:336} help to derive the transition temperature $T_{c}(h \neq 0)$ using the commonly known $T_{c}(h= 0)$ of the model in MF picture. Though both dynamics can provide the same result, to make the discussion specific, we prefer working with the Metropolis rule. For $ h =0 $, SCE~\eqref{Eq:334} yields a similar result of $\bar{m}(h =0)$ as in equilibrium  SCE~\eqref{Eq:331}.   If we consider two  arbitrary values of $\beta$ at which $ \bar{m}(h\neq 0)= \bar{m}(h =0)$, it follows that
 $\tanh[\beta J z |\bar{m}|(1+h)] = \tanh[\beta^{0} J z|\bar{m}|]$. This MFA gives 
 $(1+h)\beta(\textsc{MF}) = \beta^{0}(\textsc{MF})$ and 
 \begin{equation}
T(\textsc{MF}) = (1+h)T^{0}(\textsc{MF}).
\end{equation}
In particular, and remarkably, $\bar{m}(h\neq 0)$ and $\bar{m}(h=0)$ are equal at critical point, $\bar{m}(T_c , h\neq 0)= \bar{m}(T_{c}, h=0)$. Consequently, it becomes 
\begin{equation}\label{Eq:338}
T_{c}(\textsc{MF}) =  zJ (1+h).
\end{equation}
Here, it is straightforward to determine  $T_{c}^{0}(\textsc{MF})= Jz$ where $z=2d$ for a given $d$. For example, $T_{c}^{0}(\textsc{MF})$ is equal to $4$  for $d=2$,  and $ 6$ for $d=3$. Results obtained from numerical solutions for $z=4$ and $z=6$ are shown in Figure~\ref{fig:302} 
\begin{figure}[hbpt]
\centering
\includegraphics[width=0.52\columnwidth]{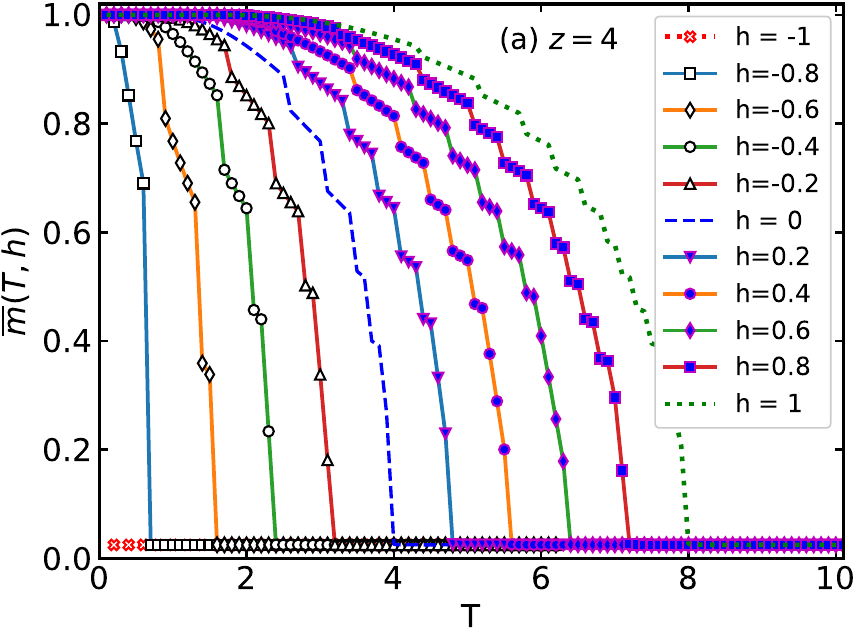}
\includegraphics[width=0.46\columnwidth]{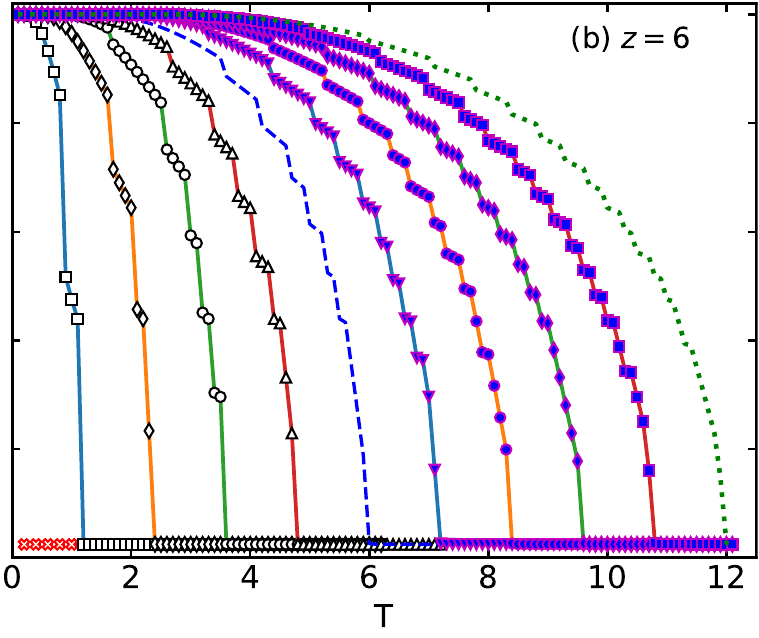}
\caption{\small{MF solution of magnetization per site $\bar{m}$ versus temperature $T$ for different values of $h$ (see keys). A calculation of the Metropolis dynamics~\eqref{Eq:334} for $z=2d$, (a) $d=2$ and (b) $d=3$. The dashed blue line ($h=0$) agrees with $T_{c}^{0}=z$. $T_{c}= 2 z$ is shown by dotted line ($h=1$).}} \label{fig:302}
\end{figure}
which are plots of the Metropolis-MF magnetization in SCE~\eqref{Eq:334} versus the temperature ($\bar{m}$ vs $T$) for various values of $h$. For $h=0$, the MF $T_{c}^{0}=z$ as we expect. For $h>0$,  $T_{c}$  increases with increasing $h$ and approaches $2z$  as $h \rightarrow 1 $. For $h < 0$, $T_{c}$ decreases with decreasing $h$ where $T_{c} \to 0$ as $h\rightarrow - 1 $. Specifically, at the critical temperature, the figure verifies the results in Equation~\ref{Eq:338} that $\bar{m}(T_{c}, h) \approx 0$ for all $h$. Therefore, the MFA for the model with given $z$ is described as follows: For $T >T_{c}\equiv (1+h)zJ$, the system is paramagnetic (PM) with $\bar{m} = 0$ and, for $T < T_{c}$, the system is ferromagnetic (FM) with $\bar{m} \neq 0$. This interpretation is qualitatively consistent with Section~\ref{S22} (see Figure~\ref{fig1}).

\newpage
\section{Results and Discussion}\label{S4_Results}
We demonstrate the results of the MC simulation for various values of the parameter $h$ using the modified update rules (\verb"B"\ref{Eq:213}) and (\verb"B"\ref{Eq:214}), as introduced in Section~\ref{S2_Methods}. In the following, we briefly describe the formalism of the required physical quantities in their usual forms. Consequently, the practice of FSS analysis follows the same fashion. Then, we present some examples of the obtained numerical results and discuss these in detail.
 \subsection{Measurement of physical quantities}
Relevant physical quantities such as the magnetization per site $m$, susceptibility $\chi $, average energy per spin  $\langle E \rangle /N $ and the specific-heat $C$ are defined as;
 \begin{eqnarray}
m & =  \langle M \rangle/N, \mathrm{\; Equation~\ref{Eq:207}}, \label{Eq:439a}\\
\chi & =  \left.\left(\langle M^{2}\rangle\frac{}{}-\langle M \rangle^{2} \right)\right/ NT ,\label{Eq:439b}\\
\langle E \rangle  / N & =   \langle E \rangle  \mathrm{ \;  in \; Equation~\ref{Eq:204} \;  per  \;  spin } ,   \label{Eq:439c}\\
C & =    \left.\left( \langle E^{2} \rangle\frac{}{}- \langle E \rangle^{2} \right)\right/ NT^{2} . \label{Eq:439d}
\end{eqnarray}
Here $N = L^{2}$ is the total number of spins, and the symbol $\langle \dots \rangle $ refers to an average in the steady state at a given $T = T(h)$. Accordingly, we look in to the (qualitative and quantitative) dependency of the macroscopic quantities on the parameter $h$.
Employing the two modified update rules, we perform MC simulations of model  \verb"B" and measure the required macroscopic quantities relevant to the investigation of PTs. To indicate the location of the transition temperature $T_{c}(h)$, we apply FSS analysis for the finite-size MC data of the Binder-cumulate $U_{4}$ associated with the distribution of the magnetization,
\begin{equation}\label{Eq:440}
 U_{4} = 1 - \langle M^{4}\rangle/(3\langle M^{2}\rangle^{2}).
\end{equation}
Further, the FSS analysis is convenient to examine the dependency of these quantities on $N$. For the usual equilibrium Ising model (and model \verb"A"), the FSS is well-known. 
Continuous PTs are classified by their critical exponents, which characterize the behavior near/at the transition point. The most relevant include: $m \sim |T-T_{c}|^{\beta}$, $\chi \sim |T-T_{c}|^{-\gamma}$,  $\xi \sim |T-T_{c}|^{-\nu }$, and $C \sim |T-T_{c}|^{-\alpha}$ where  $\beta $, $\gamma$, $\nu$, $\alpha$ are critical exponents for magnetization, susceptibility, correlation length, and specific heat, respectively. For square lattice Ising (model \verb"A"), the critical exponents $\beta = 1/8$, $\gamma = 7/4$, $\nu = 1$, and $\alpha = 0$ are known exactly. (Do not confuse the exponent $\beta$ with  $1/T$. In Sections~\ref{S4_Results} and~\ref{S5_Summary} $\beta$ refers to the critical exponent.)
In the case of model \verb"B", we treat the FSS of the steady state quantities analogous to that of model \verb"A": 
\begin{eqnarray}
m & \approx & \mathcal{M}(1-T/T_{c}) L^{(1-\beta)/\nu} \Longleftrightarrow mL^{\beta/\nu}  \textsl{ vs }  -\tau L^{1/\nu} \label{Eq:441a}   \\ 
\chi &  \approx  & -\mathcal{X}(1-T/T_{c}) L^{(1+\gamma)/\nu}\Longleftrightarrow  \chi L^{-\gamma/\nu}    \textsl{ vs }   \tau L^{1/\nu}  \label{Eq:441b} \\ 
C &  \approx  & -\mathcal{C}(1-T/T_{c}) L^{(1+\alpha)/\nu}\Longleftrightarrow  C L^{-\alpha/\nu}    \textsl{ vs }   \tau L^{1/\nu},\label{Eq:441c}
\end{eqnarray}
where $\mathcal{M}$, $\mathcal{X}$,   $\mathcal{C}$ are their respective scaling functions and $\tau = (T - T_{c})/T_{c}$ has been used here with $\tau  <  0$, $\tau  =  0$ and $\tau  >  0$ for $T < T_c$, $T= T_c$ and $T > T_c$, respectively. 
\subsection{Detailed numerical results}
We now present numerical results and  FSS analysis for model \verb"B" where $-1\le h\le 1$: (i) by varying the values of the parameter $h$ for a given system with fixed size $N = L^{2}$, and (ii) using a fixed value of $h$ for the system of various sizes. However, as introduced in Section~\ref{S22}, we consider $h=\pm 0.25$ for our detailed numerical results where the update rules (\verb"B"\ref{Eq:213}) and  (\verb"B"\ref{Eq:214}) are used for $h= 0.25$ and $h = - 0.25$, respectively.

\paragraph{Using different $h$ values for a given system:}
In the absence of an effective interaction ($h=0$), we have observed that equilibrium PT occurring at $T_{c}^{0}\approx 2.269$ as usual. Similarly, in the presence of effective interactions ($h\neq 0$), PTs also occur at $T_{c}(h)\neq T_{c}^{0}$. To begin with, we present in  Figure~\ref{fig:403} the plots of those physical quantities~(\ref{Eq:439a}$-$~\ref{Eq:439d}) as a function of $T(h)$ for various $h$ but here $L=80$ has been fixed.
\begin{figure}[hbpt]%
\centering 
$\begin{array}{cc}%
\rm{Metropolis \; (B\ref{Eq:213})} &  \rm{Glauber \;  (B\ref{Eq:214})} \\ %
\includegraphics[width=0.5 \columnwidth]{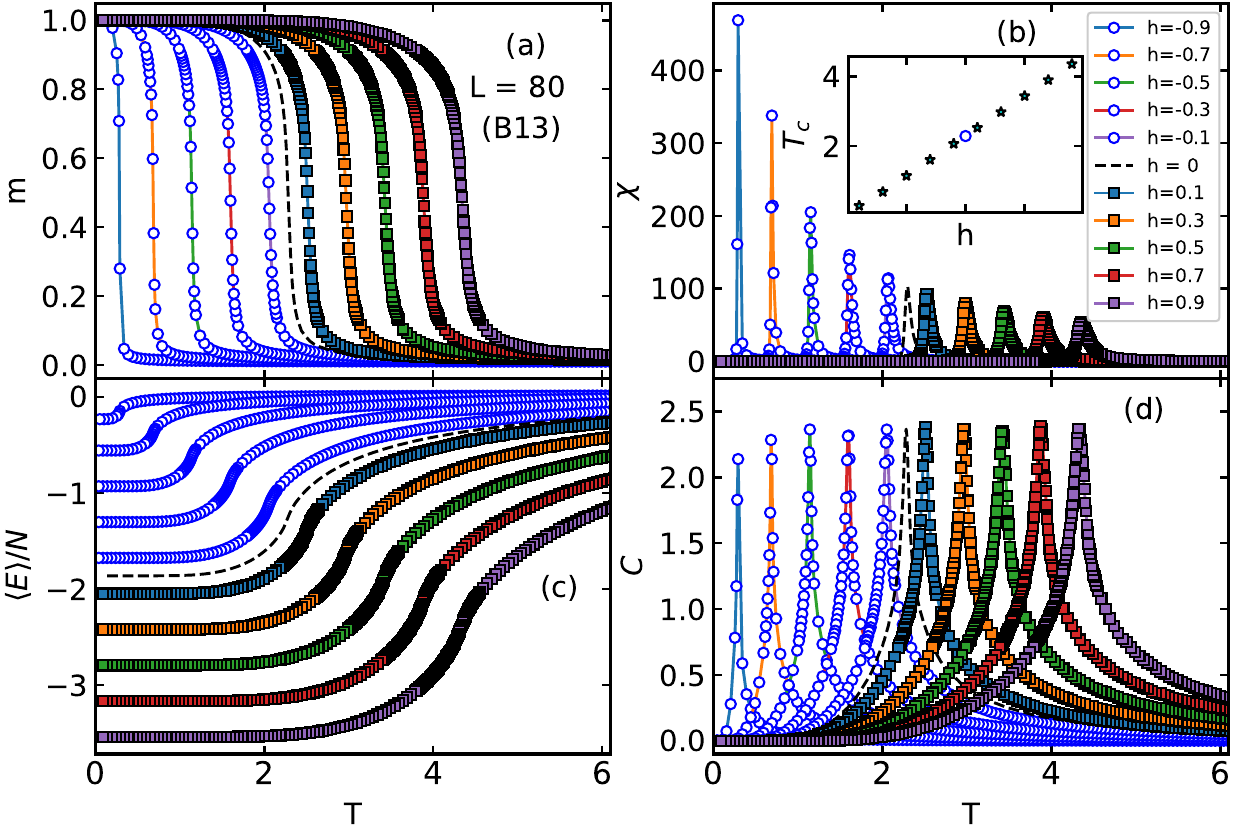} & 
\includegraphics[width=0.5 \columnwidth]{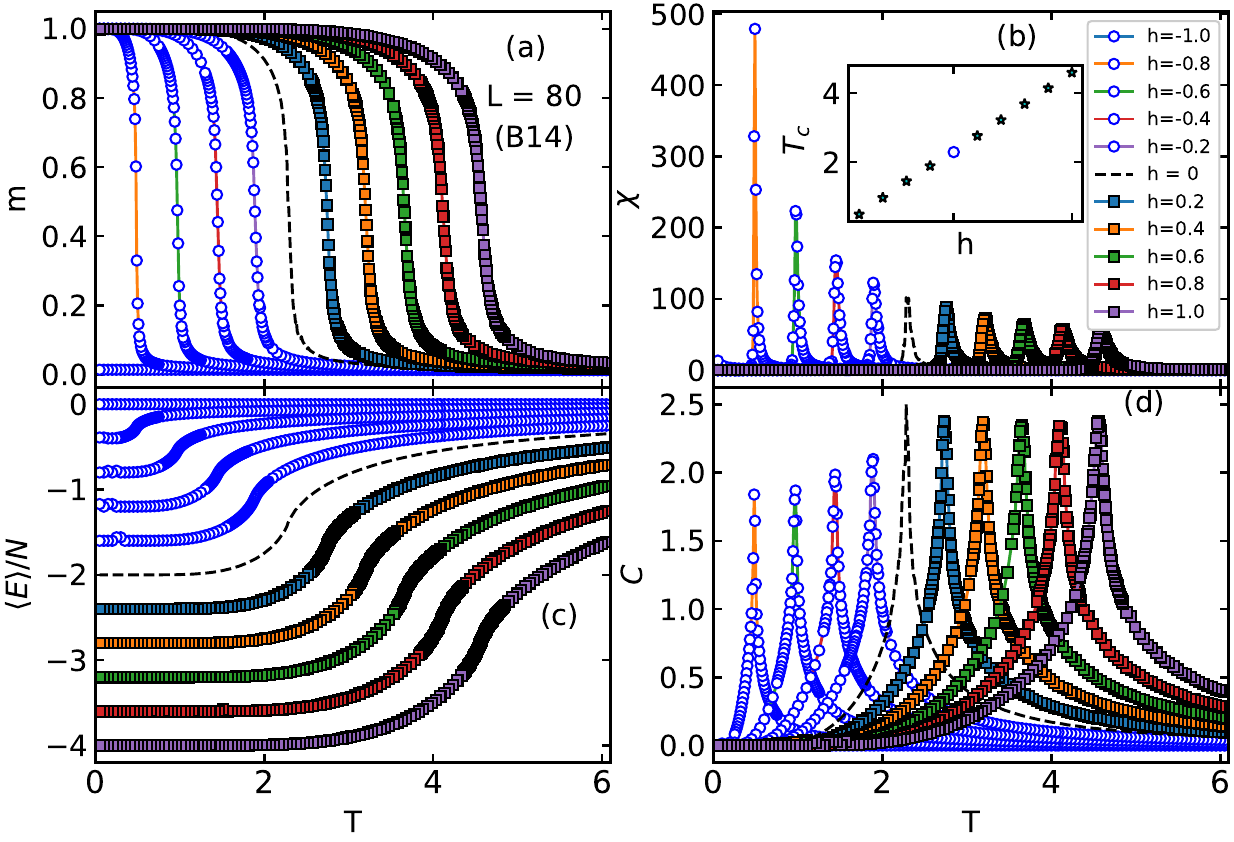}
\end{array}$%
\caption{\small{Plot of physical quantities (\textsf{a}) magnetization per spin $m$~\eqref{Eq:439a}, (\textsf{b}) magnetic susceptibility  $\chi $~\eqref{Eq:439b}, (\textsf{c}) energy per spin $\langle E \rangle $~\eqref{Eq:439c} and (\textsf{d}) specific heat $C $~\eqref{Eq:439d} as a function of temperature $T$  with varying  $h$ (see keys) but for fixed linear size $L=80$. MC results with Metropolis (B\ref{Eq:213}) and Glauber (B\ref{Eq:214}). For a given $h$, a peak position (panels \textsf{b}  and \textsf{d}) provides an estimate of $T_{c}(h)$ where the inset within panel \textsf{b} is a plot of $T_{c}(h)$ versus $h$.}} \label{fig:403}
\end{figure}
Qualitatively, they vary with varying $h$, and clearly the dashed line (`- -') retrieves the equilibrium ($h=0$) quantities as expected. The peak positions of $\chi$ versus $T$ (\textsf{b}) and that of $C$ versus $T$ (\textsf{d}) give estimates of $T_{c}(h)$. In both plots (b and d), we observe the shifting of $T_{c}$ to its relatively higher values with increasing $h$. However, the position of the peaks decreases as $h$ increases with increasing $T$ for $\chi$ vs $T$ while almost unchanged for $C$ vs $T$. 
\paragraph{Using fixed $h$ for system of various size:}
Let us proceed to analyze the critical behavior of the nonequilibrium PTs by means of FSS analysis for $h=\pm 0.25$. Figure~\ref{fig:404} exhibits plots of physical quantities~(\ref{Eq:439a}-\ref{Eq:439d}) as a function of temperature $T$ for $h= 0.25$ (\verb"B"\ref{Eq:213}) and $h=  -0.25$ (\verb"B"\ref{Eq:214}) each for $L = \{30, 40, 60, 80, 120\}$.   
\begin{figure}[hbpt]
\centering
$\begin{array}{cc}
\rm{Metropolis \; (B\ref{Eq:213})} &  \rm{Glauber \;  (B\ref{Eq:214})}\\ 
 \includegraphics[width=0.55 \columnwidth]{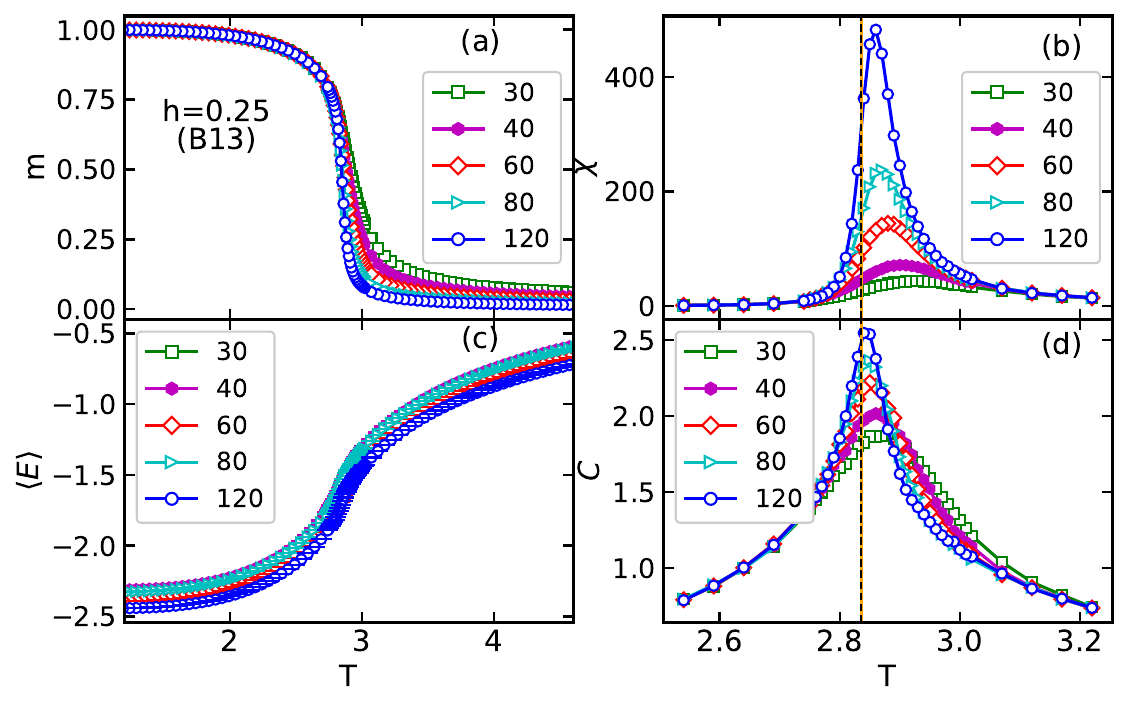}
&  \includegraphics[width=0.55\columnwidth]{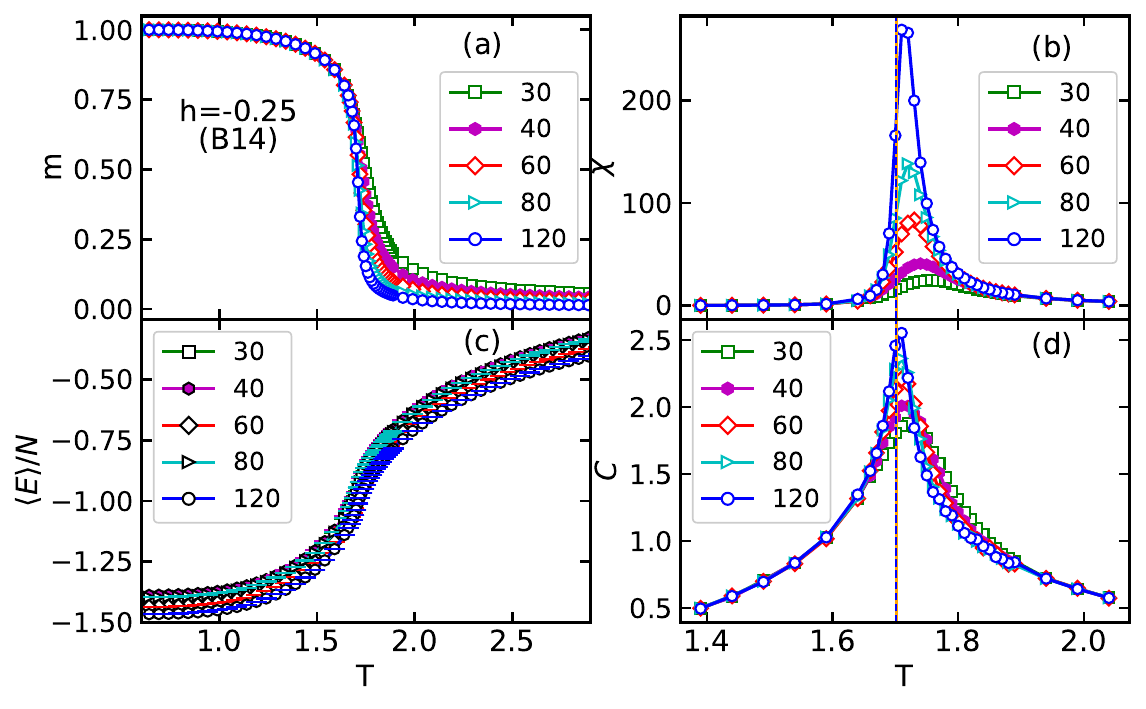}
\end{array}$
\caption{\small{Plot of physical quantities~(39) (a) $m$, (b) $\chi $, (c) $\langle E \rangle $ and (d) $C $ as a function of $T$  with varying $L$ (see keys) for each fixed parameter $h=0.25$ and $h=-0.25$. In panel \textsf{b} and \textsf{d}, the vertical (orange) line represents the location of $T_{c}$~\eqref{Eq:222}, $5/\ln(3+2\sqrt{2}) \approx 2.8365$(b) and $3/\ln(3+2\sqrt{2}) \approx 1.7019$(d).}} \label{fig:404}
\end{figure} 
In panel \textsf{b} and \textsf{d}, the vertical (orange) line represents the location of $T_{c}$~\eqref{Eq:222}, $5/\ln(3+2\sqrt{2}) \approx 2.8365$(b) and $3/\ln(3+2\sqrt{2}) \approx 1.7019$(d). With increasing $L$, the maxima $\chi^{\mathrm{max}}$ and $C^{\mathrm{max}}$ clearly grow and the locations of peak points move toward the orange lines. For both $h=\pm 0.25$, all relevant properties are discussed (see Figure~\ref{fig:405}). Figure~\ref{fig:405}(a) discusses the Binder cumulant $U_{4}$ versus $T$ for various linear sizes $L$ for $h=0.25$ (B\ref{Eq:213}) and $h=-0.25$ (B\ref{Eq:214}). From the FSS relation associated with Binder~\eqref{Eq:440}, we clearly see that $U_{4}$ for various values of $L$ converge as $T\to T_{c}$ and the curves intersect at the same point. The interaction points yield a good estimation of the transition temperatures $T_{c}=2.8357(3)$ for $h= 0.25$, and $T_{c}=1.7017(2)$ for $h=- 0.25$. 
\begin{figure}[hbpt]
\centering
$\begin{array}{ccc}
\includegraphics[width=0.32\columnwidth]{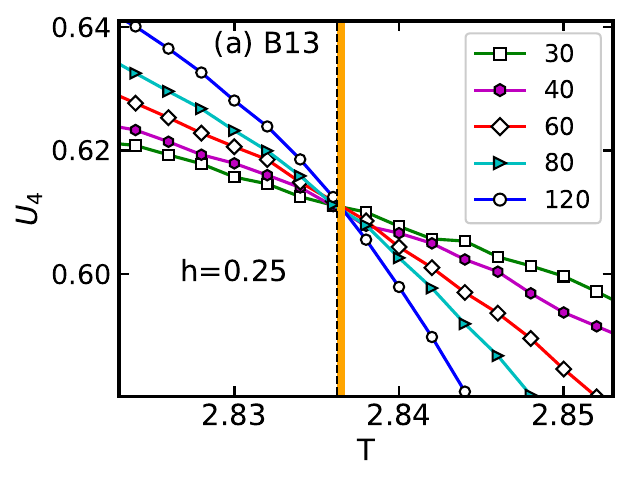}
&\includegraphics[width=0.32\columnwidth]{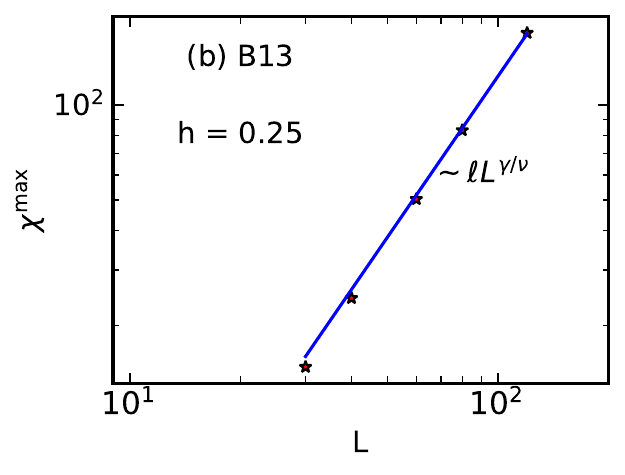} &
\includegraphics[width=0.32\columnwidth]{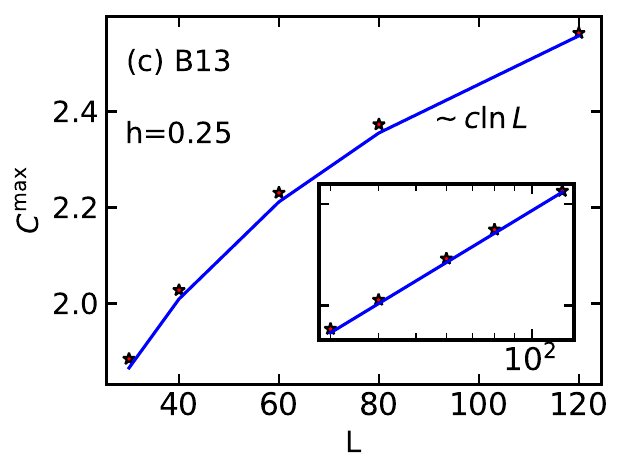} \\
 \includegraphics[width=0.32\columnwidth]{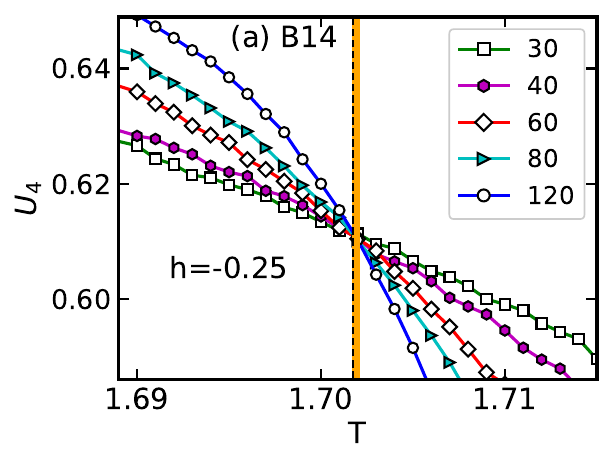} &
\includegraphics[width=0.32\columnwidth]{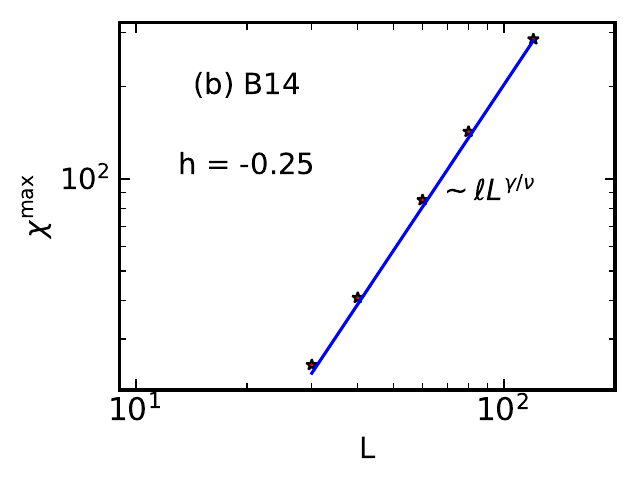} &
\includegraphics[width=0.32\columnwidth]{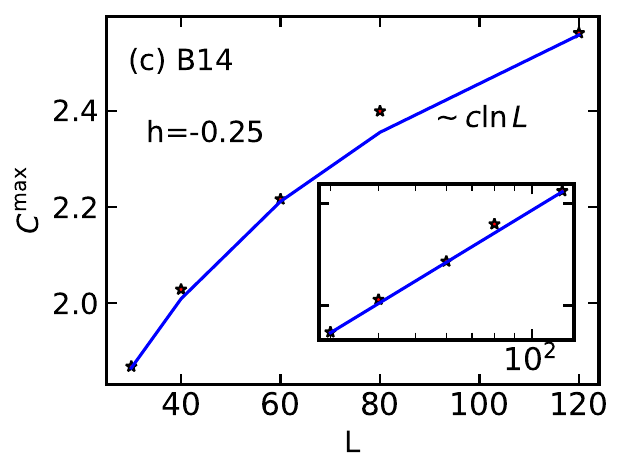}
\end{array}$
\caption{\small{(a) Binder cumulant $U_{4}$ computed for different values of $L$  plotted versus temperature $T$ for two values of  $h= 0.25$(B\ref{Eq:213}) and $h=-0.25$ (B\ref{Eq:214}).
The intersection point provides the critical temperature $T_{c}^{\mathrm{FSS}}(h)= 2.8357(3)$for $h=0.25$  and $1.7017(2)$ for  $h=-0.25$. 
(b) Plot of $\chi^{\rm{max}}(L)$ versus $L$ in log-log scale for $h=0.25$ and  $h=-0.25$.  (c)  Plot of $C^{\rm{max}}(L)$ versus $L$  for $h=0.25$ and $h=-0.25$. 
Here $\ell =c/10 $ was used to get $ \gamma/\nu \approx 1.748(2)$ for $h=0.25$ (and $ \gamma/\nu \approx 1.746(3)$ for $h=-0.25$) where the constant $c \approx 0.4995$ and $ c_{0} = c/3$. Note that the inset (c) shows semi-log scale for the same data of its linear scale, $C \sim c\ln L$.}} \label{fig:405}
\end{figure} 

Referring to Figure~\ref{fig:404} (panel b), the magnetic susceptibility  $\chi$ displays a peak of a size $L$-dependent transition  at pseudo-critical  $T_{c}^{*}(L)$. The horizontal position of the peaks shifts toward a dashed line with increasing $L$ while its vertical position increases with increasing $L$.   Likewise, the specific heat $C$ (see panel d) also displays peaks that shift in agreement with that of  $\chi$. If the FSS relation $\chi$ or $C$ in Equation~\ref{Eq:441b} and~\eqref{Eq:441c}   peaks at certain point of value $\ell_{0} $, then the peak point $T_{c}^{*}(L)$ for a given value of $L$ changes with $L$ as 
\begin{equation}\label{Eq:442}
T_{c}^{*}(L) =  \ell_{0} L^{-1/\nu}  + T_{c}^{\mathrm{FSS}}(h,\chi,C),
\end{equation}
where $\ell_{0}$ is a constant. On the other hand, the maximum value of the singular part of $\chi$ and $C$ in a finite-size system changes as 
\begin{eqnarray}\label{Eq:443a}
\chi_{L}(T = T_{c}^{*}) = \chi^{\mathrm{max}}(L) &  \sim  & L^{\gamma/\nu} \\ \label{Eq:443b}
  C_{L}(T = T_{c}^{*})  =      C^{\mathrm{max}}(L) &  \sim  &  L^{\alpha/\nu},
\end{eqnarray}
and such scenarios are shown in Figure~\ref{fig:405} (b and c).
Figure~\ref{fig:405}(b) shows plots of $\chi^{\mathrm{max}}$ versus $L$ in a double log scale in which we clearly see a linear property that agrees with the power-law form of  $\chi^{\mathrm{max}}$  given in~\eqref{Eq:443a}. The solid blue line is the best power-law fit that yields $ \gamma/\nu \approx 1.748(2)$ and $ \gamma/\nu \approx 1.746(3)$. Similarly, Figure~\ref{fig:405}(c) shows a main plot of $C^{\mathrm{max}}$ versus $L$ and an inset plot on a semi-log scale for the same data. As we clearly see from the main plot, the negative curvature in data points recognizes that $C^{\mathrm{max}}$ has a weaker power-law dependence on $L$. It has been noticed from Figure~\ref{fig:403}(d) that $C^{\mathrm{max}}$ is also not sensitive to the parameter $h$ for fixed $L$. Thus it is good evidence to consider the critical exponent $\alpha \approx 0$ in the same manner as in the equilibrium (h=0) model. As a result, its plot on a semi-log scale in the inset of the main Figure (with logarithmically scaled $L$ axis) now  distinctly shows a non-curvature  linear property in which a blue straight line is the best fit to the form,
\begin{equation}\label{Eq:444}
C^{\mathrm{max}} = c_{0} +  c \ln L,
\end{equation}
where $c_{0}\approx c/3$ is the regular part of the specific heat and $c\approx  0.4995$~\cite{gould2010statistical}. In addition, we examine the analysis as shown in Figures~\ref{fig:406} and~\ref{fig:407}.
\begin{figure}[hbpt]
\centering
$\begin{array}{c}
\includegraphics[width=0.32\columnwidth]{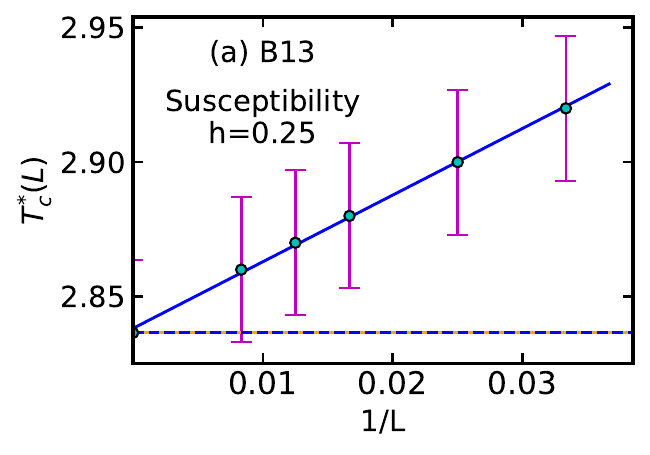} 
\includegraphics[width=0.26 \columnwidth]{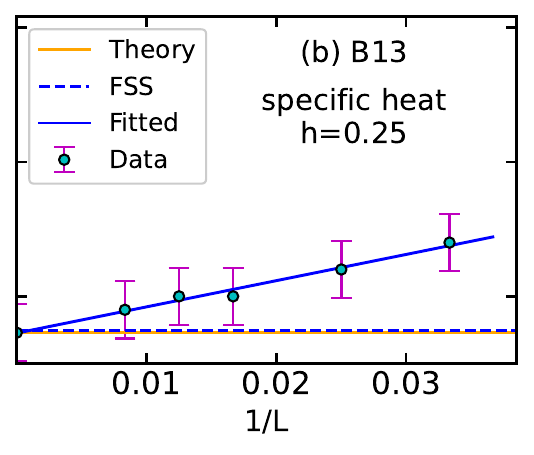} \\
 \includegraphics[width=0.32 \columnwidth]{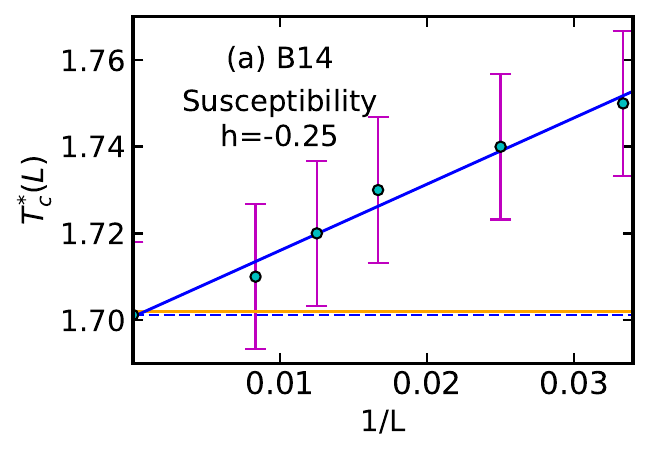}
\includegraphics[width=0.26 \columnwidth]{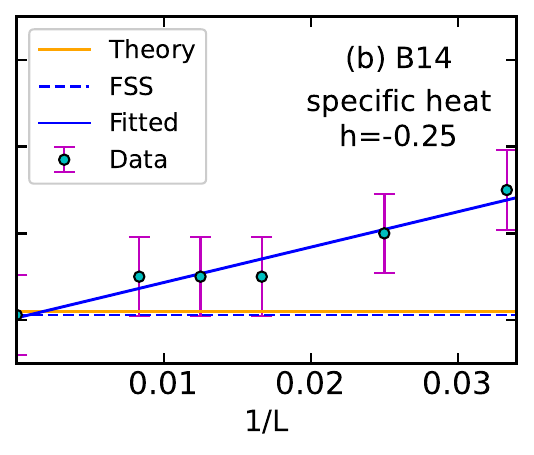} 
\end{array}$
\caption{\small{Plot of $T_{c}^{*}(L)$ versus $1/L$ where $T_{c}^{*}(L)$ refers to the \textsl{pseudo-critical} point at which $\chi$ (a) and $C$ (b) attain their maxima. Results for $h=0.25$ (B\ref{Eq:213}) and $-0.25$ (B\ref{Eq:214}). The solid blue lines are the best fitting line~\eqref{Eq:444}. The solid orange lines refer to $T_{c}$~\eqref{Eq:222} and the dashed blue lines are $T_{c}^{\mathrm{FSS}}(h=\pm 0.25)$. This yields $T_{c}^{\rm{FSS}} \approx 2.836$ (a), $2.838$ (b) and $T_{c}^{\mathrm{FSS}} \approx 1.701$ (a), $1.701$ (b).}} \label{fig:406}
\end{figure}
Figure~\ref{fig:406} shows scatter plots of the peak positions $T_{c}^{*}$ versus $1/L$ as marked by $*$ symbol. Here $T_{c}^{*}(L)$ refers to the \textsl{pseudo-critical} point at which $\chi$(\textsf{a}) and $C$(\textsf{b}) attain their maxima for $h=0.25$ (B\ref{Eq:213}) and $h=-0.25$ (B\ref{Eq:214}). The solid orange lines are the exact $T_{c}(h)$ and the dashed blue lines are $T_{c}^{\mathrm{FSS}}(h)$. The solid blue lines are the best fitting line of the form  $T_{c}^{*}(L) = \ell_{0}L^{-1/\nu} +  T_{c}^{\mathrm{FSS}}(h)$.  This yields $T_{c}^{\mathrm{FSS}}(h) \approx 2.836$ (a), $2.838$ (b) and $T_{c}^{\mathrm{FSS}}(h) \approx 1.701$ (a and b), almost reconciled  with $T_{c} =  5/\ln(3+2\sqrt{2})$ and $T_{c} =   3/\ln(3+2\sqrt{2})$. 
Further, Figure~\ref{fig:407}%
 \begin{figure}[hbpt]
\centering
\includegraphics[width=0.85\columnwidth]{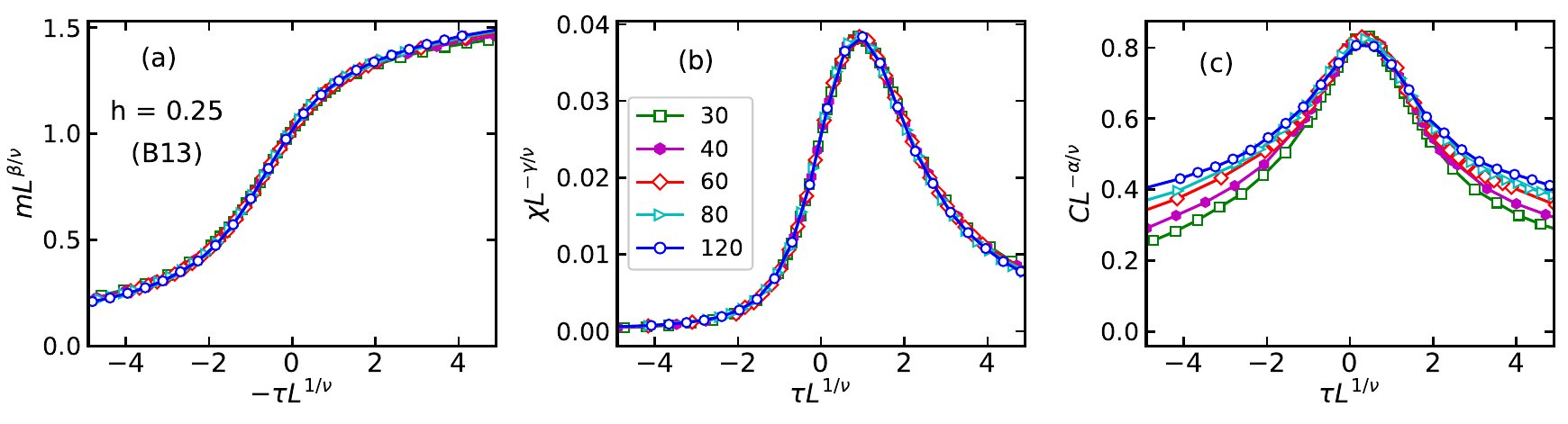}
\caption{\small{FSS of magnetization~\eqref{Eq:441a}, susceptibility~\eqref{Eq:441b} and  specific heat~\eqref{Eq:441c} for $h= 0.25$ (B\ref{Eq:213}). Plots of (a) $m L^{\beta/\nu} $ vs $-\tau L^{1/\nu}$, (b) $\chi L^{-\gamma/\nu} $ vs $\tau L^{1/\nu}$ and (c) $C L^{-\alpha/\nu} $ vs $\tau L^{1/\nu}$. The plots for the case $h= -0.25$ (B\ref{Eq:214}) are identical (not shown here).}} \label{fig:407}
\end{figure}
shows the FSS of magnetization~\eqref{Eq:441a}, susceptibility~\eqref{Eq:441b}, and specific heat~\eqref{Eq:441c} for $h= 0.25$ (B\ref{Eq:213}). An identical plot (not shown here) for $h= -0.25$ (B\ref{Eq:214}). As suggested in the FSS  of magnetization~\eqref{Eq:441a}, Figure~\ref{fig:407}(a) exhibits a plot of $m L^{\beta/\nu} $ vs $-\tau L^{1/\nu}$ from which we have systematically determined the exponent $\beta/\nu$ that provides the best scaling collapse where $\beta/\nu = 0.127(1)$ for $h=0.25$ (B\ref{Eq:213}) and $\beta/\nu = 0.125(1)$ for $h=-0.25$ (B\ref{Eq:214}). Here we use the value of $T_{c}=T_{c}^{\mathrm{FSS}}$  as estimated above. Similarly, the FSS  of susceptibility~\eqref{Eq:441b} and that of specific heat~\eqref{Eq:441c} are presented in Figure~\ref{fig:407}(b) and~\ref{fig:407}(c), respectively. The data collapses are excellent. The values of exponents that are used for $h= \pm 0.25$ are summarized  in Table~
ef{T1}.

\section{Summary}\label{S5_Summary}
This study mainly investigated NESS PTs under effective interactions in a 2D ferromagnetic Ising model on a square lattice using the Monte Carlo method. Using the modified update rules (\verb"B"\ref{Eq:213}) and (\verb"B"\ref{Eq:214}), we have performed extensive MC simulations for finite systems of various lattice sizes. We measured physical quantities such as the average magnetization per site~\eqref{Eq:439a}, magnetic susceptibility~\eqref{Eq:439b}, average energy per spin~\eqref{Eq:439c}, specific-heat~\eqref{Eq:439d}, and the Binder cumulant~\eqref{Eq:440}, for various values of $h$ in general, and two values of $h=\pm 0.25$ in particular. Table~\ref{T1} summarizes the transition temperature $T_{c}$ and the relevant critical exponents (determined using FSS techniques). We see from this table that the FSS estimations of $T_{c}$ for two $h=\pm 0.25$ values are in agreement with the analytical results~\eqref{Eq:222}. The numerical results for the exponents $\beta/\nu $ and $\gamma/\nu $ are identical to  the analytical values of the equilibrium ($h=0$) Ising model, where $\beta/\nu =  1/8 $ and $\gamma/\nu = 7/4$.
\begin{table}[hbpt]
\centering
\caption{\small{Example of critical temperature $T_{c}(h )$  for $h  = \pm 0.25$ obtained by FSS analysis of the data measured via the modified update rules (B\ref{Eq:213}) and (B\ref{Eq:214}) compared with model A~\eqref{Eq:222} and the critical exponents $\beta/\nu $ and $\gamma/\nu $. The numbers in the standard bracket denote error estimates in the last digit.}}\label{T1}
\begin{tabular}{c}  \hline 
\begin{tabular}{cccc} 
Parameter    & \; \; Critical  Temperature $T_{c}$ \;  \; &  Critical Exponent &   \end{tabular}  \\ 
\begin{tabular}{cccccc}   
$h$& \small{Analytical} & \small{Numerical} &$\beta/\nu$  & $\gamma/\nu$ & Model \\ \hline \hline
$-\infty < h < \infty $ & Equation~\ref{Eq:222} &(\verb"B"\ref{Eq:213}) and (\verb"B"\ref{Eq:214}) & $1/8$ & $7/4$ &\verb"A" \\ \hline
0.25  &$ 5/\ln(3 + 2\sqrt{2})$&$2.8357(3)$&0.127(1)&1.748(2)&(\verb"B"\ref{Eq:213}) \\ \hline 
-0.25 & $ 3/\ln(3+ 2\sqrt{2})$&$1.7017(2)$&0.125(1)&1.746(3)&(\verb"B"\ref{Eq:214}) \\  
\hline \hline 
\end{tabular}
\end{tabular}
\end{table}\\ %
In conclusion, it has shown that the numerical result of $T_{c}$ is consistent with Equation~\ref{Eq:222} as long as $-1\le h\le 1$. The obtained values of the critical exponents show that the numerical results of the scaling relations are in excellent agreement with the analytical results of the equilibrium Ising model and thus belong to the same universality class. As will be a subject of future work,  extending this investigation to the dynamic case would be helpful. Specifically, the critical behavior of the pertinent model will examined using the Langevin equation (instead of ME) and developing field-theoretic methods for the solution with the generation of long-range interactions and effects of dynamical anisotropies.

An interesting question in this context is whether this numerical method can be applicable to quickly calculate the effect of strain on 2D material critical temperature. We propose that the methods particularly used to estimate the effect of strain on the critical temperature of monolayer CrSBr and gain insights into its magnetic properties under different mechanical conditions in which the parameter $h$ could appear to represent the {\em strain levels}. MC simulations for various strain levels may allow for the characterization of strain-dependent properties of the magnetic material. Therefore, further investigation in this direction would be helpful. 
\subsection*{Abbreviations}
\begin{tabular}{@{}ll}
DB(DBC) & Detailed Balance (DB Condition)    \\
FSS  &Finite Size Scaling  \\
MC & Monte Carlo \\
ME & Master Equation \\
MF (MFA) &   Mean Field (MF Approximation) \\
NESS & Nonequilibrium Steady States \\
PT(s) & 	Phase Transition(s) \\
SCE & Self Consistency Equation\\
\end{tabular}

\subsection*{Acknowledgements}
The authors would like to thank The International Science Program, Uppsala University, Uppsala, Sweden for the support in providing the facilities of Computational and Statistical Physics lab. 
Both authors would like to acknowledge Prof. Chandan Dasgupta for his scientific comments and substantial contributions to the interpretation of data. DW thanks AAU and DDU for the support during this research work.

\subsection*{Data availability statement}
The data cannot be made publicly available upon publication because they are not available in a format that is sufficiently accessible or reusable by other researchers. The data that support
the findings of this study are available upon reasonable request from the authors.

\appendix
\section{Appendix (Supplementary Page)} \label{A:A1}
\subsection*{The modified Metropolis update rule (B\ref{Eq:213})}
According to the Metropolis update rule for the usual Ising model at temperature $T^{0}$, the standard form of the transition rate can be written as
\begin{equation}\label{Eq:A01}
W =  \texttt{MIN} \left[1, \texttt{e}^{-\Delta E^{0}/T^{0}}\right],
\end{equation}
where $W$ is a rate of transition from a state $\texttt{b}$  to  other state $\texttt{a}$,
$ \Delta E^{0} = E_{\rm{a}}^{0} - E_{\rm{b}}^{0}$, is the change in energy due to this transition.
Applications of this update rule, which satisfies DB at temperature $T^{0}$, generate equilibrium configurations of the Ising model at temperature $T^{0}$.
We want to deliberately violate the DBC to cause the system to go out of equilibrium in which the system shows a disorder-order transition which may not similar to the usual equilibrium PT. Assuming   $ \varepsilon \neq 0 $ denotes a parameter violating the DBC, $\Delta E^{0} $ in~Equation~
ef{Eq:A01} is substituted by
\begin{equation}\label{Eq:A02} 
\Delta E = \Delta E^{0} + \varepsilon\mathrm{.} 
\end{equation}
From Equation~
ef{Eq:A02}, it follows that $\Delta E > \Delta E^{0} $ for positive $ \varepsilon$  and $\Delta E < \Delta E^{0}$ for negative $ \varepsilon$. Here we notice that the former doesn't promote the flipping of spins while the latter becomes highly probable for spins to flip. Compared to spins with the usual Metropolis, spins under these flipping rates efficiently experience different temperatures. For $\varepsilon < 0$ ($  \varepsilon  > 0$), the spins may be assumed as being coupled to a thermal bath at a higher(lower) effective temperature ($T_{\rm{eff}}$) and this $T_{\rm{eff}}$   is not the same for all the spins in the system~\cite{Kumar2020}. Accordingly, this system is not in equilibrium and the PT would be a property of NESS of the system. Based on the modified  Metropolis algorithm, the transition rate for flipping a spin  $ S_{i}^{\rm{b}}\rightarrow S_{i}^{\rm{a}} $ is
\begin{equation}\label{Eq:A04}
 W(\pm S_{i} \rightarrow \mp  S_{i}) =					\left\{
  \begin{array}{ll}
    \texttt{e}^{-(  \varepsilon  \pm \Delta E^{0})/T}, & \varepsilon  \pm \Delta E^{0}  > 0 ; \\
    1, & \hbox{otherwise,}
  \end{array}
\right.
\end{equation}
where $\Delta E^{0} =  \{-8,-4, 0, 4, 8\}$. This algorithm~\eqref{Eq:A04} still \emph{respects} the DBC for $| \varepsilon | \geq 8 $. From~Equation~
ef{Eq:209}, here the ratio $\mathcal{R}(\varepsilon)$ becomes $\mathcal{R}(\varepsilon) =  \exp[-(\Delta E^{0} + \varepsilon)/T]$. For $ \varepsilon  \geq 8$, $ \mathcal{R}(\varepsilon) =    \exp[-2\Delta E^{0}/T]$ and DBC is satisfied at an effective temperature $T_{\mathrm{eff}} = T/2$, therefore, the critical temperatures at which an \emph{equilibrium} transition takes place is given by $T_c( \varepsilon  \geq 8) = 2 T_{c}^{0}$, where $T_{c}^{0} \equiv T_{c}( \varepsilon=0)  = 2/\ln(1+\sqrt{2})$~\cite{Onsager1944}. However, this algorithm~\eqref{Eq:A04} \emph{violates} the DBC for $-8 <  \varepsilon < 8$ (with $ \varepsilon \neq 0$). 
It was previously studied\cite{Kumar2020} (see also \cite{tola2023machine}) that the PT takes place in which  $T_{c}$ satisfies, 
\begin{equation}\label{Eq:A05}
T_c = \left\{
  \begin{array}{ll}
   0 < T_{c} <  T_{c}^{0}     & \hbox{for } -8 < \varepsilon  < 0; \\
  T_{c}^{0} < T_{c} < 2 T_{c}^{0} & \hbox{for \; \;  }  0 < \varepsilon  < 8 .
  \end{array}
\right.
\end{equation}
Explicitly, for $\varepsilon$ values within the range $-8 < \varepsilon < 8$,  $T_{c}( \varepsilon, \Delta E^{0})$  can be understood from the following argument.  With positive  $\Delta E^{0} = \{4, 8\}$, the ratio reads
 \begin{equation}\label{Eq:A06}
\mathcal{R}(\varepsilon) =  \exp[-(\Delta E^{0} + \varepsilon)/T].
 \end{equation}
Mapping (at a temperature or critical temperature) the known $\mathcal{R}(\varepsilon=0)$ with this~\eqref{Eq:A06}, we get $ \Delta E^{0}/T_{c}^{0} = (\Delta E^{0} +  \varepsilon)/T_{c}$. Repeat this for negative $\Delta E^{0} = \{-8, -4\}$, therefore
\begin{equation}\label{Eq:A07}
 T_{c}(h, \Delta E^{0})  = \left\{
  \begin{array}{ll}
   (1 +\varepsilon/\Delta E^{0})T_{c}^{0} & \hbox{if } \Delta E^{0}  >  0; \\
  (1 - \varepsilon/|\Delta E^{0}|)T_{c}^{0}  & \hbox{if \;} \Delta E^{0}  < 0.
  \end{array}
\right.
\end{equation}
This~Equation~
ef{Eq:A07} permits one to relate $T$ (nonequilibrium model) to $T^{0}$ (equilibrium model).  Since this relation is different for different values of $\Delta E^{0}$, it is not possible to map the probability distribution in the NESS of this model to the equilibrium distributions at a unique temperature $T^{0}$. Let us express 
 \begin{equation}\label{Eq:A08}
h_{i} = \varepsilon/|\Delta E_{i}^{0}|, \; \;  -8 \le \varepsilon \le 8 \rm{ , \; \;  and \; \;  }-1 \le h_{i}\le 1.
\end{equation}
Now it is important to compare~\eqref{Eq:A02} and~Equation~
ef{Eq:215} as long as $h(h_i)$ is defined according to~\eqref{Eq:A08}. As a result, the argument in~\eqref{Eq:A05} is almost similar to that of~Equation~
ef{Eq:225}.
Assuming that $\Delta E_{i}^{0} = \{4, 8\}$, it is quite obvious from~Equation~
ef{Eq:A07} that $T_{i} = (1+h_{i})T^{0}$. A similar result can be found for $\{-4,- 8\}$ that $T_{i} = (1-h_{i})T^{0}$.   Therefore, the modified update rule considered here violates DB at temperature $T$, and it does satisfy DB at effective temperature 
\begin{equation}\label{Eq:A09}
T_{\rm{eff}}=  T/(1\pm h)
\end{equation}
where $h:=h_{\rm{eff}}$ is the effective value. It follows that $h = (h_{1} + h_{2})/2$ where $h_{1} = \varepsilon/4 $ and $h_{2} = \varepsilon/8$. For example, given $\varepsilon = 2$, we get $h_{1}=1/2$, $h_{2}=1/4$ and $h= 3/8$ or 0.375 and for $\varepsilon = -2$, $h= -0.375$. Given $\varepsilon = \pm 4/3$, the effective parameter becomes $h= \pm 0.25$. If $\varepsilon = \pm 8$, $h= \pm 1$. Equation~\ref{Eq:A09} reads that $T_{\rm{eff}}  =  \infty$, $T_{\rm{eff}}  = T^{0}$ and $T_{\rm{eff}}  = T/2$; for $h \simeq -1$, $h =0 $ and $h=1$, respectively. The system undergoes a PT when $T_{\rm{eff}}= T_{c}^{0}$, i.e. at critical temperature $T_c = (1\pm h)T_{c}^{0}$, see Figure~\ref{fig1} and Table~\ref{TA1}. The numerical results (Section~\ref{S4_Results}) are in agreement with this conclusion.
\begin{table}[hbpt]
\centering
\caption{\small{More detail of Table~
ef{T1}.}}\label{TA1}
\begin{tabular}{c}  \hline 
\begin{tabular}{ccccc}    & \; \; Critical  Temperature $T_{c}$ \;  \; &  &Critical Exponent &   \end{tabular}  \\ 
\begin{tabular}{cccccc}   
$h$& \small{Analytical} & \small{Numerical} &$\beta/\nu$  & $\gamma/\nu$ & Model \\ \hline \hline
$-\infty < h < \infty $ & Equation~
ef{Eq:222} & (\verb"B"\ref{Eq:213}) and (\verb"B"\ref{Eq:214})  & $1/8$ & $7/4$ &\verb"A" \\ \hline
0 & $2/\ln(1+\sqrt{2})$&$-$& 0.125&1.75 & (known)\\ \hline 
$-1\le h\le 1$ & $(1\pm h)T_{c}^{0}$ & (this) &  >> & >> &\verb"A"\&\verb"B" \\ \hline
$h\ge 1$ & 2$T_{c}^{0}$ & >> & >> & >> & \verb"B" \\ \hline 
$h \le -1$ &        0    &>> &  >> & >>  & \verb"B" \\ \hline
0.25  &$ 5/\ln(3 + 2\sqrt{2})$&$2.8357(3)$&0.127(1)&1.748(2)&(\verb"B"\ref{Eq:213}) \\ \hline 
-0.25 & $ 3/\ln(3+ 2\sqrt{2})$&$1.7017(2)$&0.125(1)&1.746(3)&(\verb"B"\ref{Eq:214}) \\  
\hline \hline 
\end{tabular}
\end{tabular}
\end{table}

\end{document}